\documentstyle[aaspp]{article}

\begin{document}

\def\MSUN{\rm M_{\odot}}
\def\RSUN{\rm R_{\odot}} 
\def\MSUNYR{\rm M_{\odot}\,yr^{-1}}
\def\MDOT{\dot{M}}

\newbox\grsign \setbox\grsign=\hbox{$>$} \newdimen\grdimen \grdimen=\ht\grsign
\newbox\simlessbox \newbox\simgreatbox
\setbox\simgreatbox=\hbox{\raise.5ex\hbox{$>$}\llap
     {\lower.5ex\hbox{$\sim$}}}\ht1=\grdimen\dp1=0pt
\setbox\simlessbox=\hbox{\raise.5ex\hbox{$<$}\llap
     {\lower.5ex\hbox{$\sim$}}}\ht2=\grdimen\dp2=0pt
\def\simgreat{\mathrel{\copy\simgreatbox}}
\def\simless{\mathrel{\copy\simlessbox}}

\title{ Accretion of low angular momentum material onto black holes:
2D~magnetohydrodynamical case.
}

\vspace{1.cm}
\author{ Daniel Proga and Mitchell C. Begelman$^{1}$}
\vspace{.5cm}
\affil{JILA, University of Colorado, Boulder, CO 80309-0440, USA;
proga@colorado.edu, mitch@jila.colorado.edu}

$^1$ also Department of Astrophysical and Planetary Sciences, University
of Colorado at Boulder

\begin{abstract}
We report on the second phase of our study of slightly rotating accretion 
flows onto black holes. We consider magnetohydrodynamical (MHD) accretion 
flows with a spherically symmetric density distribution at the outer 
boundary, but with  spherical symmetry broken by the introduction of
a small, latitude-dependent angular momentum and a weak radial magnetic 
field. We study accretion flows by means of numerical 2D, axisymmetric, 
MHD simulations with and without resistive heating. Our  main result is 
that the properties of the accretion flow depend mostly on an equatorial 
accretion torus which is made of the material 
that has too much angular momentum to be accreted directly.  
The torus accretes, however, because of 
the transport of angular momentum due to 
the magnetorotational instability (MRI).
Initially, accretion is dominated by the polar funnel, 
as in the hydrodynamic inviscid case, where material has zero or very low 
angular momentum. At the later phase of the evolution,
the torus thickens towards the poles and develops a corona 
or an outflow or both. Consequently, the mass accretion through the funnel 
is stopped. The accretion of rotating gas through the torus is significantly 
reduced compared to the accretion of non-rotating gas (i.e., the Bondi
rate). It is also much smaller than the accretion rate in the inviscid,
weakly rotating case.
Our results do not change if we switch on or off resistive heating.
Overall our simulations are very similar to those presented by Stone, Pringle,
Hawley and Balbus despite different initial and outer boundary conditions. 
Thus, we confirm that  MRI is very robust and controls the nature of 
radiatively inefficient accretion flows.

\end{abstract}

\section{Introduction}

Accretion onto supermassive black holes (SMBHs) very likely powers some of 
the most dramatic phenomena of astrophysics, such as quasars and  powerful 
radio galaxies. However, SMBH accretion does not always result
in high radiative output, as evidenced by SMBHs that appear to spend
most of their time in  a remarkably quiescent state
(e.g., Di Matteo et al. 1999, 2000, 2001; Loewenstein et al. 2001;
for Sgr~A$^\ast$, see review by Melia \& Falcke 2001). 
Dim SMBHs are not 
something one would expect, because these black holes 
are embedded in the relatively
dense environments of galactic nuclei. Therefore it is natural to suppose 
that the gravity due to an SMBH will draw in matter  at high rates, 
leading to a high system luminosity.

Estimates for the accretion luminosity, $L$, rely on
assumptions about the mass accretion rate, $\MDOT_a$, and the efficiency
of transforming the gas energy into radiation, $\eta$ (i.e.,
$L= \eta c^2 \MDOT_a$).
Both $\MDOT_a$ and $\eta$ are uncertain and there is no
generally accepted model which could explain low luminosity SMBHs by
predicting low enough $\MDOT_a$ or $\eta$, or both.

The result that SMBHs are dimmer than they should be is primarily
due to the fact that we estimate $\MDOT_a$ based
on the density and temperature of the gas in which the SMBHs are embedded.
It is customary to adopt the analytic formula due to Bondi (1952)
to estimate the mass accretion rate.  Bondi (1952)
considered spherically symmetric accretion from 
a non-rotating polytropic gas with uniform density $\rho_\infty$
and sound speed $c_\infty$ at infinity. Under these assumptions, 
a steady state solution to the equations 
of mass and momentum conservation exists 
with a mass accretion rate of
\begin{equation}
\MDOT_B= \lambda 4 \pi R^2_B \rho_\infty c_\infty,
\end{equation} 
where $\lambda$ is a dimensionless parameter
that, for the Newtonian potential, depends only on the adiabatic index. 
The Bondi radius, $R_B$, is defined as
\begin{equation}
R_B=\frac{G M}{c^2_\infty},
\end{equation}
where $G$ is the gravitational constant and $M$ is the mass of the accretor.

Relatively high $\MDOT_a$ predicted by the Bondi formula
is partially responsible  for generating a lot of interest in
accretion flows with low $\eta$, that is, where transfer
of the flow internal energy to radiation is very
inefficient. Nonradiatve accretion flow solutions are possible
because binding energy dissipated in the gas can be
advected through the event horizon before being radiated
(Ichimaru 1977; Rees et al. 1982; Narayan \& Yi 1994, 1995; 
Abramowicz et al. 1995). 

Neglect of radiative cooling in accretion flows does not lead to just one
family of solutions, however. In particular, the above-mentioned
pure advection-dominated inflows constitute only one possible 
class of solutions. Once rotation is allowed, radiatively inefficient 
hydrodynamical (HD) flows become subject to strong convection 
(Begelman \& Meier 1982; Narayan \& Yi 1995), 
which can fundamentally change the flow pattern and its radiative properties 
(Igumenshchev \& Abramowicz 1999; Blandford \& Begelman 1999; 
Stone, Pringle \& Begelman 1999;
Quataert \& Narayan 1999; Narayan, Igumenshchev \& Abramowicz 2000; 
Quataert \& Gruzinov 2000). Numerical and theoretical studies show
that convection alters the steep ($\propto r^{-3/2}$) density profile of 
advection-dominated flows into a much flatter ($\propto r^{-1/2}$) profile, 
which can explain the faintness of many SMBHs 
because it predicts relatively low density close to the black hole (i.e.,
$\MDOT_a$ is low in eq.~1). Similar structural changes occur in 
the magnetohydrodynamical (MHD) limit (Stone \& Pringle 2001, SP01 hereafter; 
Hawley, Balbus, \& Stone 2001; Machida, Matsumoto \& Mineshige 2001; 
Igumenshchev \& Narayan 2002; Hawley \& Balbus 2002), 
although here the turbulence is probably
driven by magnetorotational instability  (MRI) rather than thermal convection
(Balbus \& Hawley 2002; but see Abramowicz et al. 2002 and 
Narayan et al. 2002 for alternative views).

The turbulent character of both HD and MHD models does not
settle the issue of what happens to the energy and angular momentum
that must be transported away. There are two possibilities:
(i) turbulent transport effectively shuts off the accretion flow,
turning it into a closed circulation (Narayan et al. 2000; Quataert \&
Gruzinov 2000) or (ii) turbulent transport drives powerful outflows
that can strongly modify the black hole's environment 
(Narayan \& Yi 1994, 1995; Blandford \& Begelman 1999).
Recent MHD simulations bring new insights that may help us
to resolve this issue. For example, Hawley \& Balbus's (2002) 
three-dimensional MHD simulations show that, with and without resistive
heating, some mass and energy in  nonradiative accretion flows are carried
off by an outflow in keeping with the outline of the second
possibility.

An important element of the problem of very low SMBH luminosity  
is the rate at which mass is captured into the accretion flow. 
If this rate is far lower than $\MDOT_B$, then the problem of very low SMBH
luminosity becomes less severe.
SMBHs draw matter from 
an extended medium and most authors assume that the Bondi (1952) formula 
provides 
an adequate approximation for the rate of mass supply. 
Evolution of the flow  already captured by an SMBH has been studied
extensively, but evolution of the flow including regions
beyond the domination of gravity  has not been given its due.
The Bondi formula has been derived under 
the assumption that this gas is non-rotating and only under the influence of 
the central gravity. Thus, for a given gravitational field,
the gas internal energy determines the accretion rate. By relaxing
this assumption, introducing additional forces or sources of energy, 
one may find that the mass supply rate is much lower than the one predicted by
the Bondi formula. For example, 
the rate at which matter is captured by a black
hole can be severely limited when the matter is heated by X-rays produced
near the black hole (Ostriker et al. 1976) or by  mass outflow
from the central region (Di Matteo et al. 2003; Fabbiano et al. 2003). 
In these two cases, the gas internal energy
is increased. Introducing kinetic energy to the gas at infinity
may have a similar effect:  
although the flow outside the Bondi accretion radius often can be described
as nonrotating, even a tiny amount of angular momentum, $l$ --- when followed
inward --- could severely limit the rate at which matter is captured by the
black hole (e.g., Proga \& Begelman 2003, PB03 hereafter, 
and references therein). 

In PB03, we 
reported on the first phase of our study of slightly rotating accretion 
flows onto black holes. We considered inviscid, hydrodynamic
accretion flows with 
a spherically symmetric density distribution at the outer boundary,
but with  spherical symmetry broken by the introduction of
a small, latitude-dependent angular momentum. Namely
we assumed that at the outer radial boundary, 
the specific angular momentum, $l$,  depends on the polar angle, $\theta$, as 
\begin{equation}
l(\theta)=l_0 f(\theta),
\end{equation}
where  $f=1$ on the equator ($\theta=90^\circ$) and monotonically
decreases to zero at the poles ($\theta=0^\circ$ and $180^\circ$).
PB03's  main result was that the properties of 
the accretion flow do not depend as much on the outer boundary conditions 
(i.e., the amount as well as distribution of the angular momentum
parameterized by $l_0$ and the form of $f$) as on 
the geometry of the non-accreting matter. 
The material that has too much 
specific angular momentum to be accreted 
($l>2 R_S c$, 
where $R_S=2 G M/c^2$ is the radius of a Schwarzschild black hole)
forms a thick torus near the equator. 
Consequently, the geometry of the polar region, where material is accreted 
(the funnel), and the mass accretion rate through it are constrained by 
the size and shape  of the torus. 
PB03's results showed one way in which
the mass accretion 
rate of slightly rotating gas can be significantly reduced compared to 
the accretion of non-rotating gas (i.e., the Bondi rate), and set the stage
for calculations that will take into account the transport of angular
momentum and energy as presented here.

We report on the second phase of our study and assess 
the gross properties of rotating accretion 
flows onto  black holes with the inclusion of  MHD effects. 
We consider a classic Bondi accretion flow  modified by
the introduction of (i) a small, 
latitude-dependent angular momentum at infinity, 
(ii) a pseudo-Newtonian gravitational potential and (iii)
weak magnetic field. 
The imposed angular momentum and magnetic field are weak enough  
to have initially a negligible
effect on the density distribution at the outer boundary, which remains
spherically symmetric. Contrary to PB03, we now consider the transport
of energy and angular momentum by including magnetic field effects.
Therefore our model allows for accretion of matter which initially
has a specific angular momentum  higher than $2 R_S c$.
Yet we still  consider a simple model of an accretion flow, 
simpler than those occurring 
in nature, as we neglect the gravitational field due to the host galaxy,
radiative heating and cooling effects and 3D effects of MHD. 

Our work is complementary to some other previous studies. Several authors 
considered  accretion onto black holes with a focus on the evolution of 
rotationally supported thick tori including the transport of angular
momentum and energy (e.g., Igumenshchev \& Abramowicz 1999; SP01; 
Machida et al. 2001; Hawley \& Balbus 2002; McKinney \& Gammie 2002;
Igumenshchev, Narayan \& Abramowicz 2003). The main difference between 
our work and these studies is that our simulations have one degree of
freedom more than the previous studies. The other authors consider cases 
where if not for the transport of angular momentum due to internal
stresses, there would be neither time evolution nor  mass accretion. 
For example, for their initial conditions Stone et al. (1999),  
SP01, Hawley \& Balbus (2002), and Igumenshchev et al. (2003) 
adopted a bounded torus in hydrostatic equilibrium 
with uniform angular momentum, embedded in zero angular momentum ambient gas
which is also in hydrostatic equilibrium. The zero-$l$ ambient gas
has a very low density and is unimportant dynamically.
Thus previous studies consider accretion from a finite reservoir of gas
that is not refilled during  simulations.
On the other hand, we allow time evolution and mass accretion, 
even without internal stresses. We  start our simulations 
from a radial inflow and allow for gas with a constant density
but a range of angular momenta to enter 
the computational domain during simulations.
Thus, our flow  is a complex convolution of rotating 
and non-rotating flows with similar densities 
that may be sub-Keplerian over a very large range of radii. 
We find that despite these differences 
the accretion through the torus, facilitated by MRI, dominates the inner flow.
Thus, our results reinforce the findings of the previous studies.

We note that if the low-$l$ material were allowed to accrete
onto the black hole then it could significantly  contribute 
to the total mass accretion rate.
However, detailed simulations are required to check what will
happen in such a situation. Here we report on quite intriguing
results, especially when compared to the HD inviscid
simulations: the low-$l$ material that can be accreted in 
the inviscid case (with no need for any transport of $l$)
is {\it not} accreted  in the MHD case, whereas the high-$l$ material
that could not be accreted in the inviscid case, is accreted in the MHD
case. This seemingly paradoxical behavior
can be understood by the fact that the energy and angular momentum
from the MHD torus are deposited outside the torus in the polar region
(see, e.g., Stone et al. 1999; Blandford \& Begelman 1999; 
Blandford \& Begelman 2002a, 2002b; Hawley \& Balbus 2002), effectively
choking off the funnel.
In PB03, we speculated that  this energy and angular momentum
might interfere with the inflow in the funnel. But it was unclear
whether this would lead to a significant net reduction of $\MDOT_a$,
especially when the compensating inflow in the torus was taken
into account. The simulations we present here suggest that the compensation
is negligible, and that the total accretion rate is far lower than in
the inviscid case with a similar angular momentum distribution at infinity.

The outline of this paper is as follows. We describe our calculations 
in Section 2. In Section 3, we present our results. We summarize our results 
and discuss them together with their limitations in Section 4. 

\section{Method}

\subsection{Equations}

To calculate the structure and evolution of an accreting flow, we solve 
the equations of magnetohydrodynamics
\begin{equation}
   \frac{D\rho}{Dt} + \rho \nabla \cdot {\bf v} = 0,
\end{equation}
\begin{equation}
   \rho \frac{D{\bf v}}{Dt} = - \nabla P - \rho \nabla \Phi+ \frac{1}{4\pi} {\bf (\nabla \times B) \times B},
\end{equation}
\begin{equation}
   \rho \frac{D}{Dt}\left(\frac{e}{ \rho}\right) = -P \nabla \cdot {\bf v}+\eta_r\bf{J}^2,
\end{equation}
\begin{equation}\label{eqn:induction}
{\partial{\bf B}\over\partial t} = {\bf\nabla\times}({\bf v\times B-
\eta_{\it r} J}),
\end{equation}
where $\rho$ is the mass density, $P$ is the gas pressure, 
${\bf v}$ is the fluid velocity, $e$ is the internal energy density,
$\Phi$ is the gravitational potential, $\bf B$ is the magnetic field vector,
$\bf J$ is the current density, and $\eta_r$ is an anomalous resistivity.
We adopt an adiabatic equation of state
$P~=~(\gamma-1)e$, and consider models with $\gamma=5/3$.
Our calculations are performed in spherical polar coordinates
$(r,\theta,\phi)$. We assume axial symmetry about the rotational axis
of the accretion flow ($\theta=0^\circ$ and $180^\circ$). 

We present simulations  using 
the pseudo-Newtonian potential $\Phi$ 
introduced by Paczy\'{n}ski \& Wiita (1980) 
\begin{equation}
\Phi=-\frac{G M}{r-R_S}.
\end{equation}
This potential approximates general relativistic effects
in the inner regions, for a nonrotating black hole. 
In particular, the Paczy\'{n}ski--Wiita (WP) potential
reproduces the last stable circular orbit at $r=3 R_S$
as well as  the marginally bound orbit at $r=2 R_S$.

To compute  resistivity, we follow SP01:
\begin{equation}
\eta_r=Q (\Delta x)^2 |\bf J|/\sqrt{\rho},
\end{equation}
where $\Delta x$ is the grid spacing and $Q$ is a dimensionless constant.
We also follow SP01 in adopting $Q=0.1$. 

\subsection{Initial conditions and boundary conditions}

For the initial conditions of the fluid variables we follow PB03 and
adopt a Bondi accretion flow with zero
angular momentum everywhere except for the outermost part of the flow.
In particular, 
we adopt $v_\theta=0$ while $v_r$ and $\rho$
are computed using the Bernoulli function and mass accretion rate
for  spherically symmetric Bondi accretion with the PW potential.
We set $\rho_\infty=1$ and specify $c_\infty$ through
$R'_S\equiv R_S/R_B$ (note that $R'_S=2c^2_\infty/c^2)$.
We specify  the initial conditions by adopting a non-zero
specific angular momentum $l$ for the outer subsonic part of the flow.

We consider a general case where the angular momentum at the outer radius $r_o$
depends on the polar angle  via
\begin{equation}
l(r_o,\theta)=l_0 f(\theta),
\end{equation}
with $f=1$ on the equator ($\theta=90^\circ$) and $f=0$
at the poles ($\theta=0^\circ$ and $180^\circ$). 
We express the angular momentum
on the equator  as
\begin{equation}
l_0=\sqrt{R'_C} R_B c_\infty,
\end{equation}
where $R'_C$ is the ``circularization radius'' on the equator in units of $R_B$
for the Newtonian potential (i.e.,  $ GM/r^2= v^2_\phi/r$ at $r= R'_C R_B$).

We adopt one form for the function $f(\theta)$: 
\begin{equation}
f_3(\theta)=\left\{ \begin{array}{ll}
                 0    & {\rm
                  for}~~\,~~\theta~~<~~\theta_o~~{\rm and}~~\theta >
                 180^\circ -\theta_o  ~~\,~~\\
                 l_0  & {\rm for}~~\,~~\theta_o\le~~\theta~~\le 
                   180^\circ-\theta_o.
\end{array}
\right.
\end{equation}
We call this function $f_3$ to be consistent with the nomenclature in PB03.

We generate the initial magnetic field using a vector potential,
i.e., $\bf B=\nabla \times A$.
We consider one straightforward initial magnetic configuration:
a purely radial field defined by the potential 
${\bf A} = (A_r=0, A_\theta=0, A_\phi= A  \cos{\theta}/r \sin{\theta})$.
We scale the magnitude of the magnetic field
using a parameter, $\beta_o \equiv 8 \pi P_B(r_o)/B^2$
defined as the plasma parameter $\beta\equiv 8\pi P/ B^2$ at 
the outer boundary, $r_o$, so that
\begin{equation}
A ={\rm sign}(\cos{\theta})\sqrt{(8 \pi P_B(r_o)/\beta_o)} r_o^2,
\end{equation}
where $P_B$ is the gas pressure associated with the Bondi solution
at $r_o$. Note that the magnetic field changes sign across the equator.

Our standard computational domain is defined to occupy the radial range
$r_i~=~1.5~R_S \leq r \leq \ r_o~=~ 1.2~R_B$ and
the angular range
$0^\circ \leq \theta \leq 180^\circ$. We consider models with $R'_S=10^{-3}$.
The $r-\theta$ domain is discretized into zones with 140 zones in 
the $r$ direction and 100 zones in the $\theta$ direction.
We fix zone size
ratios, $dr_{k+1}/dr_{k}=1.05$, and
$d\theta_{l}/d\theta_{l+1} =1.0$ for
$0^\circ \le \theta \le 180^\circ$. However, we have also performed
some runs with 
$d\theta_{l}/d\theta_{l+1} =1.02$ for $0^\circ \le \theta \le 90^\circ$ 
and $d\theta_{l+1}/d\theta_{l} =1.02$ for $90^\circ \le \theta \le
180^\circ$ 
(i.e., the zone spacing is decreasing toward the equator).

The boundary conditions are specified as follows. At the poles,
(i.e., $\theta=0^\circ$ and $180^\circ$), we apply an axis-of-symmetry 
boundary condition. At both the inner and outer radial boundaries, 
we apply an outflow boundary condition for all dynamical variables 
except the magnetic field.  
For the magnetic field at the outer boundary, we apply an outflow condition,
whereas at 
the inner boundary, 
we follow SP01 and  use a negative stress condition (i.e., we
enforce $B_r B_\phi \leq 0$ at $r=r_i$). We also ran some models using
an outflow condition for the magnetic field at the inner boundary
and found similar results.
As in PB03, to represent steady conditions at the outer radial
boundary, during the evolution of each model we continue to apply the
constraints that in the last zone in the radial direction, $v_\theta=0$,
$v_\phi=l_0 f(\theta)/ r \sin{\theta}$, and the density is fixed at the 
Bondi value at all times. Note that we allow $v_r$ to float.
Additionally, we fix the magnetic field at its initial
radial configuration in the last zone in the radial direction
(i.e., 
$B_r=\frac{1}{r\sin{\theta}} \frac{\partial(A_\phi \sin{\theta})}{\partial \theta}$,
$B_\theta=0$ and $B_\phi$ is allowed
to float).
To reduce the problems caused by very high 
Alfv${\acute{\rm e}}$nic velocities 
in regions of very low density
(i.e., to prevent the time step from being prohibitively small),
we set a lower limit to the density on 
the grid  as $\rho_{min}(r)=\sqrt{r_i/r}$ and enforce it at all times in all models.

To solve eqs. (4)-(7) we use the ZEUS-2D code described by Stone \& Norman 
(1992a, 1992b), modified  to implement the PW potential and resistive
heating.

\section{Results}

We specify our model by several parameters.
We set the length scale in terms of 
the black hole radius in units of the Bondi radius, $R'_S$. 
Our second parameter is the adiabatic index, $\gamma$.
The third parameter (or a function rather) is the angular
momentum at the outer radial boundary, $l=l_0 f(\theta)$.
The last parameter is the plasma parameter at the outer boundary, $\beta_o$.

For practical reasons, we consider a relatively large value of $R'_S=10^{-3}$
(see PB03 and below for reasons why $R'_S < 10^{-3}$ is not suitable 
for our purposes). 
Our choice of $R'_S$ allows us to run our models over a couple of
dynamical time scales at large radii and therefore obtain 
solutions which have lost memory of the initial conditions.
We consider flows with $\gamma=5/3$. Note that we allow the entropy
of gas to increase due to nonadiabatic  heating caused by the artificial
viscosity and the resistivity (The artificial viscosity
is a compressible bulk viscosity, not a shear viscosity,
and does not affect rotation of the flow [Stone \& Norman 1992a]).
We assume an angular momentum distribution at the outer radial
boundary  as  described in Section~2. We focus our attention on accretion 
of matter with low angular momentum, i.e., 
where the corresponding centrifugal force  is small compared
to  gravity  for all $\theta$ at the Bondi radius (see PB03 for details).
Finally, we consider weak magnetic fields only. Our choice for 
the magnetic field strength is dictated by the requirement that
the flow be initially super-Alfv${\acute{\rm e}}$nic in the entire computational domain:
$|v_p|> |v_{Ap}|$, for all radii,
where $v_p\equiv \sqrt{v^2_r+v^2_\theta}$ 
is the poloidal fluid velocity and 
$v_{Ap}\equiv \sqrt{(B_r^2+B^2_\theta)/4\pi\rho}$ is
the poloidal Alfv${\acute{\rm e}}$n velocity.

Table~1 summarizes the properties of the simulations we discuss
here. Columns (2) through (7) give the numerical resolution in 
the radial direction;
the black hole radius compared to the Bondi radius, $R'_S$;
the circularization radius compared to the Bondi radius, $R'_C$;  
the specific angular momentum on the equator at $r=r_o$, $l_0$, in
units of $2 R_S c$; 
the width of the angular distribution for which $l \le 2 R_S c$, $\theta_o$;
and the angular momentum dependence on the polar angle at the outer
boundary, $f(\theta)$,  respectively. Columns (8) and (9) give the
plasma parameter at the outer boundary, $\beta_o$, and the dimensionless
constant of the anomalous resistivity, $Q$, respectively.
Table~1 also presents the final time at which we stopped each simulation
(all times here are in units of the Keplerian orbital time at $r=R_B$),
the range within which
the maximum specific angular momentum at the inner radial boundary
varies at the end of the simulation, 
$l_{a}^{max}$, and 
the mass accretion rate through the inner radial boundary
measured near the end of the simulation,  in units of  the corresponding 
Bondi accretion rate. The mass accretion rate is time-averaged
over 0.2 orbits at the end of each simulation, except for run~C
where it is averaged for 0.05 orbits only. 
Finally, column (13) gives comments about runs different
from the standard runs (e.g., higher spatial resolution
in the $\theta$ direction near the equator).

Our simulations show that for $l_0\ge 2R_S c$ 
the accretion flow consists of an equatorial torus with 
MHD turbulence driven by  MRI. The latter produces
accretion through the torus. The MHD turbulent torus also determines
the fate of the material in the polar funnel,  
where $l< 2R_S c$  and could be accreted directly. For 
cases with a very weak magnetic field
or more generally at the beginning of simulations,
there is  supersonic funnel accretion as in the  inviscid HD case.
However, as MRI in the torus grows, the torus thickens and an outflow
from the torus develops. As a result, the torus or its wind closes
the polar funnel and quenches the accretion through the  funnel. 
We describe an example of such an accretion flow in some detail first
(Section 3.1). This is followed by a limited parameter survey in
which we focus on varying two key aspects of our models:
the strength of the magnetic field and
the angular distribution of angular momentum on 
the outer boundary.

\subsection{Accretion flow consisting of an MHD torus}

In this section we describe the properties
and behavior of our model in which $R'_S=10^{-3}$, 
$\beta_o=10^6$, and
a step function describes  the angular distribution of angular momentum on 
the outer boundary (run~D). 
We assume that for  $45^\circ \le \theta \le 135^\circ$, 
the specific angular momentum on the equator at 
the outer boundary equals  $2 R_S c$, 
whereas for $\theta < 45^\circ$ and $\theta > 135^\circ$, $l=0$.
This angular distribution of $l$ on 
the outer boundary uses the same $\theta_o$ as the fiducial
model in PB03 (i.e., model B04f1a) for which $R'_C=0.1$ and 
$f(\theta)=1-|\cos{\theta}|$. 
Thus we assume that
the material that cannot be accreted onto the black hole, without
angular momentum transport, is located (at the outer boundary)  relatively
close to the equator.  Inversely, we consider  a relatively wide polar
funnel containing zero-$l$ material. Our run~H is the HD inviscid
counterpart of run~D and serves as a reference run.

Figure~1 presents a sequence of density and angular velocity
contours, and the direction of the fluid velocity for run~D.
After a transient episode of infall, the gas with $l= 2 R_S c$  
piles up outside the black hole and forms a thick torus
bounded by a centrifugal barrier near the rotation axis. 
Soon after the torus forms (i.e., a couple of orbits at $r=r_i$), 
the magnetic field is amplified both by MRI and by shear. 
The torus starts evolving rapidly and accretes onto 
the black hole. Another important effect of magnetic fields is
that the torus produces a magnetized corona and an outflow. 
By a corona we mean gas outside a torus, 
with a low $\beta$ ($\simless 0.1$),  density lower than in the torus 
by one or more orders of magnitude, and 
with vigorous circulation (see the top panels in Figure~8
for a good illustration of the torus and corona).   
On the other hand, by an outflow we mean gas with 
a systematic poloidal velocity
directed away from the equator and the black hole.
Initially, the role of the corona and outflow  is negligible. 
For example,
at the early phase of the evolution, 
$0.1 \simless t \simless 0.4$, the low-$l$ material close to the axis
can accrete almost steadily through a funnel despite the torus
corona and outflow (see Fig. 4). 
The mass accretion rate during this phase is dominated by the low-$l$
material and is  similar to $\MDOT_a$
for a pure HD inviscid flow where accretion occurs only through 
the polar funnel. The accretion rate due to the torus
is $\simgreat$ 1 orders of magnitude lower than $\MDOT_a$ due to the polar
funnel.
But at the later phase, $t> 0.5$, 
the MHD turbulent corona of the torus expands toward the poles and the torus
outflow becomes stronger. They both eventually shut off 
the polar funnel accretion. Note that the polar outflow (`jet')  reaches
the outer boundary by $t \sim 0.5$.
A shock wave, associated with a transient, propagates  outward 
through the entire computational domain during the first dynamical
time scale. After the shock wave passes through the outer boundary,
the flow in the torus is subsonic, highly variable and is directed inward near
the equator and outward close to the poles. 
We note that the flow with non-zero angular momentum 
keeps rotating on cylinders over the whole simulation regardless of
how complex is the velocity field. We shall return to this point below.
Also we note that at the end of the simulation, the angular distribution
of captured mass differs
significantly from the spherically symmetric Bondi accretion flow.
Only gas which enters the computational domain near the equator
reaches the inner part of the domain. 
We will show the inner part of the flow and discuss its nature below.
However, it is clear even from Fig. 1 that some of the equatorial
inflow does not make it into the black hole but rather turns around
and leaves the computational domain through the outer boundary.

To provide some insight into the time dependence, Figures~2 and 3 show
the time evolution of the mass accretion rate in units of the corresponding
Bondi rate. Initially, $\MDOT_a$ drops from 1 to 0.03 at $t=0.1$.
Then it rises sharply by one order of magnitude and starts
oscillating irregularly 
between $\sim 0.003$ and $\sim 0.03- 0.1~\MDOT_a/\MDOT_B$ for 
the remainder of the simulation.  Figure~3a shows  that toward the end
of simulation, $\MDOT_a(t)$ has settled into a pattern that
can be characterized as follows:
$\MDOT_a$ sharply increases, then decreases gradually by $\sim$~1 order
of magnitude to an almost steady level and then increases sharply again.
These sharp increases and gradual decreases are quasi-periodic 
and reoccur about every  0.07 orbits. Figure~3a
shows also that on the top of this quite regular pattern of `bursts'
and exponential `declines' there are short-lived `dips' and 'spikes'.
In summary, the time-dependence and level of accretion via the torus 
appear to have reached a persistent state.

To show the complex structure of the flow at small radii and gain some
insight into  the very highly time-dependent evolution of the mass accretion 
rate, Figures~4--7 and Figure~8
compare the inner accretion flow at four different 
times when a) accretion occurs through the torus and polar funnel
($t=0.22$, Fig.~4 and the top left panel in Fig.~8); 
b) accretion occurs only through the torus ($t=2.39$, Fig.~5 and the top right
panel in Fig.~8);
c) accretion through the torus is  quenched by the strong magnetic
field which forms a magnetized polar cylinder around the black hole;
($t=2.41$, Fig.~6 and the bottom left panel of Fig.~8); 
and d) accretion of low-$l$ material occurs
through the polar region outside the torus ($t=2.36$, Fig.~7 and the bottom
right panel of Fig.~8).
Figures~4---7 plot snapshots of the density,  
entropy, $S\equiv \ln(P/\rho^\gamma)$, angular velocity, specific angular 
momentum, and direction of the fluid velocity (top panels from left to right) 
and the total pressure, $P_{tot}=P+B^2/8\pi$, 
magnetic pressure, $P_{mag}=B^2/8 \pi$, 
plasma parameter, $\beta$, toroidal magnetic field, and 
direction of the magnetic field (bottom panels from left to right).
To display the generic features of run~D in its
four states in a more copact form, 
Fig.~8 compares only snapshots of the density
over-plotted by the direction of the fluid velocity.
Note vertical arrows on Figure~2 and Figure~3a which mark
the times for these four snapshots: arrows a, b, c, d correspond 
to Figs.~4, 5, 6 and 7 (and  top left, top right, bottom left
and bottom right panels in Fig.~8), respectively.

Figure~4 shows the inner flow characteristic of the early phase
of the evolution ($0.1 \simless t \simless 0.4$). The torus made of 
high-$l$ material
accretes onto the black hole due to  MRI. The torus produces
a magnetized corona with vigorous circulation bounded  by the low-$l$
material accreting onto the black hole through the polar funnel.
The flow at this stage is similar to its HD inviscid 
counterpart because the polar funnel accretion is dominant
and its $\MDOT_a$ is determined by the shape of the torus.
There are two qualitative differences between run~D
in this early phase and its HD inviscid counterpart.
In run~D, the torus also accretes  but at much lower rate
than the funnel. In the HD inviscid case there is no torus accretion at all.
The shape of the funnel in run~D is
also affected by the torus corona, albeit slightly.
However, as the torus corona and outflow grow, the above mentioned 
differences become important.

Figure~5 shows the inner flow at a later phase of the evolution,
when the corona and outflow have fully developed and can block
the low-$l$ material incoming in the polar funnel. At this
point accretion is due only to the torus and $\MDOT_a$
is at an intermediate level. During this intermediate $\MDOT_a$ state, 
matter is accreted via a torus extending down to the inner boundary. 
There can also be some accretion through the polar funnel but its rate is very 
low and the actual value of the accretion rate via the funnel is determined
by the density floor we imposed (see Section~1). For small radii, 
the polar funnel is ``empty''  because the torus outflow extends to 
the poles at large radii and shuts off the mass supply. However, 
the polar funnel has a significant pressure due to the magnetic field.
In fact, the total pressure distribution is close to spherical despite 
the huge change in the gas  pressure as we go from the poles to 
the equator for a given radius. The torus is time-variable due to 
the MHD turbulence. 
In particular, the density of the accreted material fluctuates and once it 
becomes too low, the magnetic field in the polar funnel can expand toward
the equator and reconnect. At that instant the torus is pushed
outward by the magnetic field and $\MDOT_a$ drops until 
the gas in the torus piles up and squashes the magnetic field (compare
Figure~5 and 6). Fig.~6 shows that
at  very small radii, the magnetic field is vertical
in this state instead of being radial as during other 
states shown.  Each `dip' in the accretion
rate is followed by a `spike' during which the torus unloads
the material piled up during the dip.

The strong time variability of the inner accretion flow
affects also the outer accretion flow. For example, the torus
corona and outflow struggle constantly with the low-$l$
material in the polar region. This material has neither rotational
nor pressure support and therefore it would accrete if not for the
torus corona and outflow. However, the torus corona and outflow
sometimes become too weak to prevent the low-$l$ material
from accreting. When this happens we observe a burst of accretion.
Figure~7 illustrates such a burst. Note a low-$l$ stream reaching 
the inner boundary below the equator. It takes a relatively
long time ($\sim 0.03$ orbits) for the torus corona and outflow to 
push this stream of  low-$l$ material away. Therefore
$\MDOT_a$ decreases gradually with time to the level determined
by the accretion rate due to the torus alone.

We conclude that the time-dependent accretion in run~D is due to
unsteady accretion via the torus, the presence of a very strong
magnetic field at small radii, and low-$l$ material trying
to accrete outside  the torus. The tension of the magnetic field tries
to stop the gas incoming  from the torus. Once the torus accretes
too slowly the magnetic tension causes to the field lines to straighten
and seals off the black hole because 
a highly magnetized cylinder forms around it.
At this stage, gas flowing through the torus is constrained to move
roughly parallel to the polar axis. This continues until enough low-$l$
gas builds up in the torus to overcome the tension of the magnetic
cylinder.

Despite the differences in the flow in these various states,
there are many properties of the flow that stay unchanged.  
We allow the entropy
of gas to increase due to nonadiabatic  heating caused by the artificial
viscosity in run~D. However, we note that the entropy is increased
only inside the polar flow and is nearly constant inside the torus.
This property indicates that MRI, not convection, is responsible
for the complex nature of the flow. In particular,
the vigorous circulation inside the torus is driven by MRI.
The same is true when resistive heating is switched on (e.g., run~F).
In the discussion of the flow at large radii (Figure 1), we noted
that despite complex velocity and magnetic fields the flow
rotates on cylinders. Figs. 4--7 show that this property
holds even at small radii, as indicated by the contours of $\Omega$  parallel
to the rotation axis. The contours of specific angular
momentum reveal that most of the changes in $l$ occur
near the equator just outside the torus. Specifically, $l$ decreases
from 1 to 0 with decreasing radius near the equator. However, inside
the torus, $l$ remains nearly constant with clear indication
of some regions with $l> 1$! 
The very fact that we observe regions inside the torus with $l>1$ indicates 
the outward transport of $l$ because our initial and
boundary conditions do not introduce any material with $l>1$.
In passing, we note that we decided to consider a step function
for the angular distribution of $l$ because we wish to
see as clearly as possible the evolution of $l$.

We start our simulation with a very weak magnetic field and very slow
rotation. Although our flow rotates differentially, 
the initial rotation is far from Keplerian in its
values as well as its dependence on radius. The latter scales
like $r^{-2}$, typical for a constant angular momentum fluid,
rather than like $r^{-1.5}$ as for a Keplerian disk. In fact,
our initial conditions are such that  Keplerian rotation is  reached
only at $2 R_S$. Since MRI is driven by the free energy in differential 
rotation, we have analyzed our solution in great detail 
to check whether  MRI is indeed responsible for the nature of our flow.
In particular, we have checked that the flow (the torus, more precisely) is 
subject to MRI for the actual rotational profile and the strength
of the magnetic field (see eq. 108 in Balbus \& Hawley 1998) 
and that our numerical resolution is adequate to follow the growth
and saturation of MRI. Our analysis shows that most of the inner
torus is unstable and that we are able to resolve, although marginally, the 
fastest growing MRI mode inside the inner torus.
The wavelength of the mode is $\lambda_c=2 \pi v_A/\sqrt{3} \Omega$.
For example, in run~D we find that over 
most of the central region of the torus $\lambda_c/\Delta x \sim 4$
but there are also regions with $\lambda_c/\Delta x \sim 100$ 
and with $\lambda_c/\Delta x \simless 1$.

To investigate the nature of our solutions in more detail, we have 
performed some simulations with a higher resolution near the equator,
where MRI is expected to be most important. We find that
our test run~E, with
$d\theta_{l}/d\theta_{l+1} =1.02$ for $0^\circ \le \theta \le 90^\circ$ 
(i.e., the zone spacing is decreasing toward the equator in this region)
and $d\theta_{l+1}/d\theta_{l} =1.02$ for $90^\circ \le \theta \le 180^\circ$,
gives very similar results to those from run~D.
Additional confirmation of our interpretation of the results
can be found by comparing them with expected scaling relations 
and with other relevant published results.
Therefore, we will now present our results in a form
similar to that presented by SP01.
As will be clear from the following figures, 
our results resemble the other results despite
significantly different initial and outer boundary conditions.
Generally, we find that once  MRI starts to operate
it totally determines the nature of the flow. 

Figure~9 shows the radial profiles 
of several quantities in run~D, angle-averaged over a small wedge near
the equator (between $\theta=87^\circ$ and $93^\circ$), and time-averaged
over 12 data files covering a period when accretion is nearly
steady and occurs only through the torus 
(i.e., orbits $2.388$  through $2.417$; see Fig. 3a).
We can compare this plot directly with Figure~6 in SP01 for their run~F. 
We indicate the location of the last
stable circular orbit by the vertical dotted line in each panel.

As in the MHD models of SP01, the profiles of each variable are not simple
power-laws but are rather complex. In particular, for $r'\simless 0.01$
the density is nearly constant whereas for larger radii it decreases
with increasing radius almost as $r^{-1}$. However,  we note that
this power-law at large radii is a relict of the initial conditions,
i.e., the subsonic part of the Bondi inflow with a PW potential.
The gas pressure is higher than the magnetic
pressure for large radii. However, for $r'\simless 0.004$ the two pressures
are comparable. For $r'> 0.004$, the magnetic pressure decreases with
increasing radius much faster than the gas pressure. The rotational
velocity is always sub-Keplerian. However, it peaks at nearly the Keplerian
value for $r'=0.006$. 
For $r'>0.006$, the rotational velocity scales as $r'^{-2}$ 
as for the angular momentum conserving fluid. 

To measure the Reynolds stress, we follow Hawley (2000, see also SP01) 
and compute the difference between the angular momentum flux and
mass flux times the mean angular momentum:
\begin{equation}
<\rho v_r \delta v_\phi>=<\rho v_r v_\phi>-<\rho v_r><v_\phi>,
\end{equation}
and then normalize it to the gas pressure:
\begin{equation}
\alpha_{gas}=<\rho v_r \delta v_\phi>/<P>.
\end{equation}
We find that for $r'< 0.002$
the normalized Reynolds stress is negative and decreases sharply 
with decreasing radius (note that Figure~9 shows only the magnitude
of the normalized Reynolds stress). For $r'> 0.002$, 
the normalized Reynolds stress
is positive, peaks at $0.15$ for $r'\sim 0.0025$ and decreases
with increases radius, for $r'\simless 0.0045$.
Beyond   $r'\sim 0.0045$, $\alpha_{gas}$ stays positive but is
very small. On the other hand, the Maxwell stress, normalized to the
magnetic pressure,
$\alpha_{mag}\equiv <2 B_rB_\phi>/<B^2>$,  is negative
for all radii except for a small radial range around $r'=0.027$
(note a local minimum in Fig.~9).
Comparing the actual stresses rather than the `alphas'
(i.e., $<\rho v_r \delta v_\phi>$ vs. $<B_r B_\phi/2>$),
we find that the Maxwell stress transports angular momentum
outward and  is stronger than the Reynolds stress.

The last panel in Figure~9 shows that the toroidal component
of the magnetic field is dominant at all radii. However,
for $r' \simless 0.004$, the radial component increases
with decreasing radius and becomes comparable with
the toroidal field. 

Overall, we find that the properties
of the inner flow of our run~D are strikingly similar
to those found by SP01 in their run~F. The main differences occur for very 
small radii ($r'<0.004$). For example, we  find weaker advection of such
quantities as angular momentum by the infalling gas.
This is consistent with the fact that in SP01
simulations, the torus accretes almost steadily (at least at small 
radii) whereas we find a very strong poloidal magnetic field
parallel to the rotation axis, which tends to interrupt
the torus accretion at small radii.

\subsection{Dependence of accretion flow properties on $\beta_o$}

Our simulations with an initial magnetic field weaker
than that for run~D (i.e., $\beta_o> 10^6$)  show that  
the accretion flow is qualitatively insensitive to
$\beta_o$. The main difference is in the time it takes
the torus corona and outflow to push away the low-$l$ material
inflowing in the polar funnel.
In particular, our simulations for $\beta_o=10^7$ (run~G)
show that the polar funnel accretion is stopped only after $\sim 1$ 
orbit (see Fig. 2). 
Before that time, the accretion flow is similar to its inviscid HD
counterpart: accretion occurs through a polar funnel where
$l<1$. The shape of the funnel  and $\MDOT_a$ are determined 
by the shape of a torus where there is little accretion ( $\MDOT_a=0.2$ 
but the torus contribution is less than 1\%). Only when
the torus corona and outflow become well-developed do they 
shut off accretion through the funnel.
Other differences between run~G and D are: (i)
at the end of run~G, the torus accretion rate
is lower than for run~D and (ii) in run~G, the torus accretion 
is not interrupted by the magnetic field at small radii. The decrease of
$\MDOT_a$ via the torus 
with increasing $\beta$ has been found before (e.g., SP01).
We attribute the disappearance of dips in $\MDOT_a$
to the fact that the polar funnel
is more sensitive than the torus
to the  initial and outer boundary magnetic field. 
In particular, the magnetic field in the torus
is amplified both by  MRI and by shear and is less dependent
on the initial and outer conditions than the magnetic field
in the polar funnel, which is amplified by accretion of the initial
field.

For $\beta_o < 10^6$, the initial magnetic field is too strong
to be consistent with our initial requirement of weak magnetic fields
(e.g., run~C). 
The main inconsistency is
due to the fact that for $\beta_o< 10^6$ the torsional waves are faster
than the flow. Consequently, the waves reach the inner boundary before
the material with $l>0$ and the inner flow is disrupted
by the waves reflected from the inner boundary.
We observe a train of persistent shocks propagating outward
which prevent the non-zero $l$ material from reaching small radii.

We conclude that the accretion flow in run~D, which we described 
in  detail in the previous section, is a representative
solution for a range of  magnetic fields in the weak
field regime, and is not just applicable to 
one particular initial field strength.
Next we discuss our results for various $l_0$ and demonstrate
that the generic features of run~D appear to be robust and apply
to a wide range of conditions.
 
\subsection{Dependence of accretion flow properties on $l_0$}

Our choice of $l_0=1$ and a step function for the angular distribution
of $l$ in run~D  was motivated by a wish to see if even a minimal
angular momentum can reduce the mass accretion rate. We note that by setting 
$l_0=1$ we created
rather unfavorable conditions for  MRI to grow because
the rotational velocity is sub-Keplerian in the whole
computational domain except at $r=2 R_S$.  
As we showed above, MRI is very robust 
and dominates the nature of the flow even for sub-Keplerian rotation.

In reality the angular momentum distribution beyond the Bondi radius
is likely to be complex. In particular, we expect a large range of $l$.
Numerical experiments, such as ours, try to isolate the key elements
of the accretion flow. For example, Stone et al. (1999),
SP01, Hawley, Balbus \& Stone (2001), and Hawley \& Balbus (2002)
considered a constant angular momentum hydrostatic torus for their initial
conditions. Contrary to us, they assumed a larger angular momentum
so that the circularization radius is larger than the black hole
radius by a factor of few or more.

To make more direct contact with those  previous
simulations, we ran a couple of simulations with $l=4 R_S c$ 
(runs~I and J).
We expect that a higher $l$ will result in the development of
a Keplerian flow for large radii, as the circularization radius is $8 R_S$
and MRI will be needed to transfer angular momentum.

Our run~J is an HD inviscid accretion flow and serves as a reference run.
We find that run~J is consistent with our other inviscid HD accretion
runs performed for PB03. Namely, we observe
the formation of a torus with $l> 2 R_S c$, via which there is no accretion
but the shape of which determines the geometry and $\MDOT_a$ of a polar
accretion funnel. The funnel accretion and its $\MDOT_a$ settle
into a steady state after $t \sim 0.6$. The torus exhibits subsonic
circulation. The specific angular momentum in the funnel is
practically zero, as expected. The torus rotates on cylinders
and $l$ increases from $2 R_S c$ to $4 R_S c$ with increasing radius for
$r'\simless 0.1$ at the end of our simulations. Beyond 0.1, $l$ is constant.
The gradual change of $l$ inside the torus is caused by a mixing
of the zero $l$ and non-zero $l$ material during the initial phase
of the evolution. During the later phase, the material with 
$2 R_S c \le l \le 4 R_S c$ remains in the torus as it cannot be accreted.

Run~I is similar  to run~J with the exception that we added a magnetic
field ($\beta_o=10^6$).  In the early stages of run~I, the flow consists 
of an accretion funnel and a torus.
As in run~J, $l=0$ in the funnel and
the torus is made of the material with $l> 2R_S c$.
However, contrary to its inviscid counterpart, the torus in run~I
accretes onto the black hole due to MRI.
Qualitatively, run~I is similar to runs~D and G.
The main difference is in the duration of the 
phase when both the torus and polar funnel accrete (i.e., for run~I
it lasts for $\sim 2$ orbits 
while for run~D  the duration is $\sim0.3$ and for run~G 
it is $\sim 1.3$).

Figure~10 presents the radial profiles 
of several  quantities in run~I angle-averaged over a small wedge near
the equator (between $\theta=87^\circ$ and $93^\circ$), and time-averaged
over 20 data files covering a period when accretion is nearly
steady and occurs only through the torus 
(i.e., orbits 2.35  through 2.40).
We can compare this plot directly with our Figure~9 and with Figure~6 in SP01 
for their run~F. As expected, run~I is even more similar to 
SP01's run~F than it is to run~D. In particular, we find that in run~I 
the rotational velocity is close to Keplerian for quite a wide range
of radii. This is a clear indication of the outward transport of angular
momentum. We note that for small radii ($0.003\simless r'\simless 0.02$),
the density profile in run~I scales almost as $r'^{-1/2}$
and the density is lower than in run~D for the same radii. However,
we would be very cautious about drawing any conclusions from this scaling
because of the relatively small range of radii in our computational
domain (see Section~4).

Finally, we comment on our runs~B and A with zero-$l$ material which
are simply magnetized Bondi flows. We consider these runs as tests
of our code. Because these runs are for weak magnetic fields
($\beta=10^6$ and $10^5$ for run~B and A, respectively), 
they should be very similar to the Bondi flow (see PB03).
Our simulations show that the mass accretion rate
is indeed equal to the Bondi rate and the flow is almost 
spherically symmetric. The departure from spherical symmetry is due to 
numerical resistivity (in both runs) and artificial resistivity
in run~A. However, the heating caused by magnetic field
reconnection is limited  to the region very close to the equator
and is very low, i.e., the maximum increase  of entropy is
on the equator at the inner radius, $r_i$, and 
is  0.3 per cent for run~B, and 2 per cent for run~A.

\section{Discussion}

This paper presents the second phase of our study of slightly rotating 
accretion flows onto  black holes where, in contrast to our previous paper, 
we have included magnetic fields. By adding MHD effects, we can calculate 
turbulent stresses generated self-consistently by MRI and thus include
the transport of energy and angular momentum outward as needed to accrete 
matter with a specific angular momentum higher than $2 R_S c$. 
Our simulations support strongly our 
hypothesis that even very slow rotation of gas at large radii may be 
sufficient to reduce the mass accretion rate 
to the level required by observations. 
In what follows we will summarize our results,  briefly  review 
the limitations of our work, and discuss how the physical effects 
neglected here may change the results.

We have performed numerical 2D, axisymmetric, MHD simulations
of rotating accretion flows onto a black hole. Our simulations are 
complementary to previous MHD simulations which considered strongly
rotating accretion flows. We consider slightly rotating flows and 
attempt to mimic the boundary conditions of classic Bondi accretion flows 
as modified by the introduction of a small, latitude-dependent 
angular momentum at the outer boundary, a pseudo-Newtonian gravitational 
potential and weak poloidal magnetic fields. A weak radial magnetic field and 
the distribution of $l$ with latitude allow the density distribution at 
infinity to approach spherical symmetry. The main result of our simulations 
is that the nature of the accretion flow is totally controlled by magnetic 
fields when $l\simgreat 2 R_S c$. As in the HD inviscid case,
which we studied in PB03,
the material with $l\simgreat 2 R_S c$ forms an equatorial torus.
However, in the MHD  case, the torus accretes
onto the black hole because of  MRI and produces both a corona and 
an outflow.
We find that the latter two can be strong enough to prevent 
accretion of the low-$l$ material through the polar regions, the source
of accretion in the HD inviscid case. 
The net mass accretion rate through the torus is lower than the Bondi rate
and also lower than the accretion rate in the HD inviscid case. 
In PB03, we found  that in order to reduce the mass accretion rate in
the HD inviscid case, the polar accretion funnel had to be
much narrower than the  funnels allowed 
by a hydrostatic $l=2 R_S c$ torus. Here we find that
the outflow and corona produced by the accreting torus are natural
mechanisms to narrow or even totally close the polar funnel for 
the accretion of the low-$l$ material. 
We conclude that the inclusion of even slow rotational motion of 
the MHD flow at large radii can significantly
reduce $\MDOT_a$ compared to the Bondi rate.

The accreting torus is the crucial component of our accretion flow,
therefore we describe its main properties in more detail.
Overall, the inner flow consists of a turbulent, gas pressure-dominated MHD 
torus with an outflow or corona or both, bounded by 
a magnetic pressure-dominated  polar flow.
The torus accretes near the equator. 
The accretion through the torus can be supplemented, 
in a quasi-periodic manner,
by the low-$l$ material in the polar regions. In fact, $\MDOT_a$ due 
to this polar accretion can be one order of magnitude larger than
that due to the torus. The `off torus' accretion is a manifestation
of the facts that our initial and outer boundary conditions allow the zero-$l$
material to approach the black hole
and that the torus corona and outflow are not always
strong enough to push it away.
Accretion through the torus can also be interrupted
for a short time by  strong poloidal magnetic field
that builds up during accretion. However, we observe that 
even during periods when the torus is truncated, there is inflow
of material inside the torus and its mass and pressure build up.
Consequently, the magnetic field is quickly pushed
inward by the torus and  the gas from the torus can again fall 
onto the black hole. 
Thus, our simulations show a strong buildup of the poloidal
magnetic field as in Igumenshchev, Narayan \& Abramowicz (2003)
but here we find that it is a transient and not a final solution.

Hawley \& Balbus (2002) identified three
well-defined dynamical components in their simulations 
of 3D MHD accretion flows:
(i) a hot, thick, rotationally dominated Keplerian disk, (ii) 
a surrounding magnetized corona with vigorous circulation and outflow,
and (iii) a magnetically confined jet along the centrifugal funnel.
We also observe  these three components in cases where the circularization
radius is well outside the black hole, i.e., in cases similar to those
studied by Hawley \& Balbus (2002; see also SP01).
We conclude that the three-component accretion flow is robust at small radii,
despite the fact that we also allow for low-$l$ material at large radii.

We note that $R'_S$ in our simulations is much larger than that in real 
systems, where it is $10^{-5}$ and smaller. Additionally, our relatively 
large $R'_S$ and high $\gamma$ do not allow us to capture,
over a great radial range, the asymptotic behavior of the Bondi accretion
flow, i.e., the flow in free fall yielding
a steep ($\propto r^{-3/2}$) density profile.
The free fall approximation is valid for radii small compared to the
sonic point, $x_s$. However,  our choice of $R'_S$ and $\gamma$ yield
$x_s=0.027$ (PB03). 
Thus the approximation that $\rho \propto r^{-3/2}$
is valid over less than one order of magnitude in radius. This radial
range is additionally reduced at very small radii because 
the free fall velocity in the PW potential is higher than for the
Newtonian potential, so that the density profile is flatter than 
$\rho \propto r^{-3/2}$ for very small radii.
For slightly rotating flows as in real systems, we lack analytic predictions.
Therefore, we treat our simulations only as first-order indications
of what may be happening in real systems. In particular,
we find that our accretion flows differ significantly from 
advection-dominated accretion flows
(ADAF) solutions. For example, ADAF solutions predict nearly spherically
symmetric inflows with a steep density profile, 
whereas we find a thick torus with a flat ($\sim r^{-1/2}$) density profile.
Our solutions differ also from convection-dominated accretion flows
(CDAF) solutions as we find
an MHD turbulent torus with an outflow instead of a convection-driven
closed circulation. Our solutions appear to be  MHD analogs
of the HD viscid self-similar solutions of Blandford \& Begelman (1999), where
inflows as well as outflows of gas play a critical role in determining
the properties of the accretion flow.

Our simulations do not include some of the physical processes that may
be important in accretion flows onto black holes. 
Additionally, we have performed our simulations
in 2D instead of  3D. Thus, in our simulations, a poloidal 
magnetic field can eventually dissipate according to the
antidynamo theorem (e.g., Moffatt 1978). Additionally,
we can not simulate the toroidal field MRI and we may be emphasizing
the ``channel'' solution mode (Hawley \& Balbus 1992)
which produces coherent streaming in the disk plane instead
of more generic MHD turbulence. Obviously,  3D MHD simulations
are required. However, one  should not underestimate
the importance of the larger radial dynamic range and time scale afforded
by 2D simulations if one wants to make direct contact with observations 
and with theoretical models which often yield self-similar solutions.
Future work should also include studies of accretion flows with various
initial configurations of the magnetic field.

ACKNOWLEDGMENTS: We thank J.M. Stone for useful discussions.
DP acknowledges support from NASA under LTSA grant NAG5-11736 and
support provided by NASA through grant  AR-09532
from the Space Telescope Science Institute, which is operated 
by the Association of Universities for Research in Astronomy, Inc., 
under NASA contract NAS5-26555.
MB acknowledges support from NSF grant AST-9876887.

\newpage
\section*{ REFERENCES}
 \everypar=
   {\hangafter=1 \hangindent=.5in}

{
  Abramowicz, M.A., Chen, X., Kato, S., Lasota, J.-P., Regev, O. 
  1995, ApJ, 438, L37

  Abramowicz, M.A., Igumenshchev, I. V., Quataert, E., \& Narayan, R. 2002,
  ApJ, 565, 1101

  Balbus, S.A., \& Hawley, J.F. 1998, Rev. Mod. Phys., 70, 1.

  Balbus, S.A., \& Hawley, J.F. 2002, ApJ, 573, 749

  Begelman, M.C., \& Meier, D.L. 1982, ApJ, 253, 873

  Blandford, R.D., \& Begelman, M.C. 1999, MNRAS, 303, L1
 
  Blandford, R.D., \& Begelman, M.C. 2002a, in preparation

  Blandford, R.D., \& Begelman, M.C. 2002b, in preparation
  
  Bondi, H. 1952, MNRAS, 112, 195
 
  Di Matteo, T., Allen, S.W.,  Fabian, A.C., Wilson, A.S., \& Young, A.J. 2003,
  ApJ, 582, 133

  Di Matteo, T., Carilli, C.L., \& Fabian, A.C. 2001, ApJ, 547, 731

  Di Matteo, T., Fabian, A.C., Rees, M.J., Carilli, C. L., \& Ivison, R.J. 
  1999, MNRAS, 305, 492

  Di Matteo, T., Quataert, E., Allen, S.W., Narayan, R., \& Fabian, A. C.
  2000 MNRAS, 311, 507 

  Fabbiano, G.,  Elvis, M.,  Markoff, S.,  Siemiginowska, A.,  
  Pellegrini, S.,  Zezas, A., Nicastro, F.,  Trinchieri, G., 
  \& McDowell, J.,  2003, ApJ, in press (astro-ph/0301297)

  Hawley, J.F. 1992, ApJ, 528, 462
 
  Hawley, J.F., \& Balbus, S.A. 1992, ApJ, 400, 595

  Hawley, J.F., \& Balbus, S.A. 2002, ApJ, 573, 749

  Hawley, J.F., Balbus, S.A., \& Stone, J.M. 2001, ApJ, 554, L49

  Ichimaru, S. 1977, ApJ, 214, 840.

  Igumenshchev, I.V., \& Abramowicz, M.A. 1999, MNRAS, 303, 309
 
  Igumenshchev, I.V. \& Narayan, R. 2002, ApJ, 566, 137

  Igumenshchev, I.V., Narayan, R., \& Abramowicz M.A. 2003, ApJ, submitted 
(astro-ph/0301402)

  Loewenstein, M., Mushotzky, R.F., Angelini, L., Arnaud, K.A., \& 
  Quataert, E. 2001, ApJ, 555, L21

  Machida, M., Matsumoto, R., \& Mineshige, S. 2001, PASJ, 53, L1

  Melia, F. \& Falcke, H. 2001, ARA\&A, 39, 309

  McKinney, J.C. \& Gammie, C.F. 2002, ApJ, 573, 728

  Moffatt, K. 1978, Magnetic Field Generation in Electrically Conducting 
  Fluids (Cambridge: Cambridge Univ. Press)

  Narayan, R., Igumenshchev, I.V., \& Abramowicz, M.A. 2000, ApJ, 539, 798

  Narayan, R., Quataert E., Igumenshchev, I.V., \& Abramowicz, M.A. 2002,
  ApJ, 577, 295
 
  Narayan, R., \& Yi, I. 1994, ApJ, 428, L13
 
  Narayan, R., \& Yi, I. 1995, ApJ, 444, 231
  
  Ostriker, J.P., McCray, R., Weaver, R., \& Yahil, A. 1976, ApJ, 208, L61
 
  Paczy\'{n}ski, B., \& Wiita, J.P. 1980, A\&A, 88, 23

  Proga, D., \& Begelman M.C. 2003, ApJ, 582, 69 (PB03)

  Quataert, E., \& Gruzinov A. 2000, ApJ, 545, 842
 
  Quataert, E., \& Narayan R. 1999, ApJ, 520, 298

  Rees, M.J., Begelman, M.C., Blandford, R.D., \& Phinney, E.S. 1982, 
  Nature, 295, 17

  Stone, J.M., \& Norman, M.L. 1992a, ApJS, 80, 753

  Stone, J.M., \& Norman, M.L. 1992b, ApJS, 80, 791

  Stone, J.M., \& Pringle, J.E. 2001, MNRAS, 322, 461 (SP01)

  Stone, J.M., Pringle, J.E., \& Begelman, M.C. 1999, MNRAS, 310, 1002

}

\newpage

\begin{table*}
\footnotesize
\begin{center}
\caption{ Summary of parameter survey.}
\begin{tabular}{l c  c c c c c c c c c c l  } \\ \hline 
         &       &    &            &             &          &   & & &  \\
Run & Resolution   & $R'_S$ & $R'_C$  & $l_0$ & $\theta_o$ &
$f(\theta)$& $\beta_o$ & $Q $ & $t_f$     & $l^{max}_a$  & $\MDOT_a/\MDOT_B$ & Comments  \\ 
         &                &   &            &     &       & & & & &  \\

A   &   140        & $10^{-3}$    &     0       & 0 & 0 &
------        & $10^5$     & 0.1 & 0.28&  $0.0$ &   1.0 &    \\   

B &   140        & $10^{-3}$    &       0      & 0 &   0 &
      ------- & $10^6$     & 0.1 & 0.48 &  $0.0 $ &   1.0 &    \\

C &   140        & $10^{-3}$    & $8\times10^{-3}$& 1& $44^\circ$ &
step function & $10^5$     & 0.1 & 0.1&  $0.0 $ &   0.05&    \\

D  &   140        & $10^{-3}$    & $8\times10^{-3}$& 1& $44^\circ$ &
step function & $10^6$     & 0.0 & 2.54&  $0.0-0.9$ &   0.025 & fiducial run   \\   

E &   140        & $10^{-3}$    & $8\times10^{-3}$& 1& $44^\circ$ &
step function & $10^6$     & 0.0 & 0.57 &  $0.0-0.9$ &   0.058 &  
$d\theta_{l}/d\theta_{l+1} =1.02$   \\   

F &   140        & $10^{-3}$    & $8\times10^{-3}$& 1& $44^\circ$ &
step function & $10^6$     & 0.1 & 0.32 &  $0.0-0.9$ &   0.053 &    \\

G  &   140        & $10^{-3}$    & $8\times10^{-3}$& 1& $44^\circ$ &
step function & $10^7$     & 0.0 & 2.51&  $0.0-0.9$ &   0.001 &    \\   

H &   140        & $10^{-3}$    & $8\times10^{-3}$& 1& $44^\circ$ &
step function & $\infty$  & 0.0 & 0.64&  $0.9$ &   0.17 &    \\   

I  &   140        & $10^{-3}$    & $3.2\times10^{-2}$& 2& $44^\circ$ &
step function & $10^6$     & 0.1 & 2.54 &  $0.0-0.9$ &   0.013 &    \\   

J &   140        & $10^{-3}$    & $3.2\times10^{-2}$& 2& $44^\circ$ &
step function & $\infty$     & 0.1 & 0.78&  $0.0-0.9$ &   0.27 &    \\

\hline
\end{tabular}

\end{center}
\normalsize
\end{table*}

\eject 

Fig. 1. A sequence of logarithmic density and angular velocity contours
(left and middle panels, respectively)
and velocity direction plots (right panel) from run~D at times 0.11,
0.34, 0.52, 1.45, and 2.54.
The minimum and maximum of log $\rho$ are 0.25 and 2. We use eight
equally spaced contour levels for log~$\rho$.
The  angular velocity is in units of $2 c/R_S$. The minimum
of log $\Omega$ is -6 and 
seven contour levels are equally spaced at intervals of 
$\Delta {\rm log} \Omega =0.25$.
We show the direction of the velocity field by unit vectors.

Fig. 2. The time evolution of the mass accretion rate in units of the
Bondi rate, for run~D (solid line) and run~G (dashed line).
Vertical arrows mark times for run~D corresponding to the snapshots
shown in Figures 4, 5, 6, and 7 (arrow a, b, c, and d respectively)
and in Figure~8.

Fig. 3. Late time evolution of the mass accretion rate in units of the
Bondi rate, for run~D (top panel) and run~G (bottom panel).
This figure is very similar to Fig. 2 but it 
shows in more detail $\MDOT_a$ as a function of time
toward the end of the simulations.
Vertical arrows mark times corresponding to the snapshots
shown in Figures 5, 6, and 7 (arrow b, c, and d respectively)
and in Figure~8.

Fig. 4. Two-dimensional structure of various quantities 
from the fiducial model (run~D) near the beginning of simulations
at $t=0.22$, marked by arrow~a in Fig.~2. 
At this time accretion onto the black hole occurs through
both the torus and the polar funnel.
The top panels from left to right are snapshots of
log $\rho$, $S$, log $\Omega$, $l$, and  direction
of the fluid velocity. The bottom panels from left to 
right are snapshots of log $(P+B^2/8\pi)$, log $B^2/8\pi$, log $\beta$, 
log $|B_\phi|$,
and direction of the poloidal magnetic field. All contour levels
are equally spaced.
There are 
eight contours for the density, log $\rho$,  between 0.25 and 2;
         nine for the entropy, $S$,     between 40 and 44; 
          four for the angular velocity, log $\Omega$, between -2.5 and -1.75; 
        eleven for the specific angular momentum,  $l$, between 0.1 and 1.1
(the level for 1.1 is indicated by the dotted contour);
         six for the total pressure, log $(P+B^2/8\pi)$,  between 20.5 and 23; 
         ten for the magnetic pressure, log $B^2/8\pi$,  between 17.5 and 22; 
         seven for the plasma parameter, log $\beta$,  between -3 and 3
(the levels for log $\beta< 0$ are indicated by the dotted contours);  and
         six for the toroidal field, log $|B_\phi|$,  between 7.5 and 10.

Fig. 5. Two-dimensional structure of various quantities 
from the fiducial model (run~D) at $t=2.39$, marked by arrow~b in
Figs.~2 and 3a.
The snapshots are for the same quantities as in Fig. 4.
The contour levels are also as in Fig. 4. 
This figure presents an example of an inner flow where accretion
occurs only through the torus (see also Fig. 3a).

Fig. 6. Two-dimensional structure of various quantities 
from the fiducial model (run~D) at $t=2.41$, marked by arrow~c in
Figs.~2 and 3a. 
The snapshots are for the same quantities as in Fig. 4.
The contour levels are also as in Fig. 4. 
This figure presents an example of an inner flow where 
there is no  torus accretion but only very weak accretion through
a very low density magnetized polar cylinder.
This  state  is very short-lived (see Fig. 3a).

Fig. 7. Two-dimensional structure of various quantities 
from the fiducial model (run~D) at $t=2.36$, marked by arrow~d in
Figs.~2 and 3a.
The snapshots are for the same quantities as in Fig. 4.
The contour levels are also as in Fig. 4. 
This figure presents an example of an inner flow where 
accretion is dominated by low-$l$ material which managed to reach
the inner boundary despite a blocking  corona and outflow from the torus.
The torus also accretes at this time but at a lower rate
than in the low-$l$ inflow. As the low-$l$ inflow is gradually 
pushed away by the torus corona and outflow, 
$\MDOT_a$ slowly decreases to the level where accretion occurs
only via the torus.

Fig.~8. Maps of logarithmic density overplotted by the direction
of the poloidal velocity. This figure compares the inner flow
in four different accretion states (see figs.~2 and 3a) 
shown in more detail in Figs.~4--7.

Fig. 9.  Radial profiles of various quantities from run~D, 
time-averaged from $2.388$  through $2.417$ orbits. During this period
accretion occurs only through a torus, as illustrated in Fig. 5
(see also Fig. 3a).
To construct each plot, we averaged the profiles
over angle between $\theta=87^\circ$ and $93^\circ$.
The top middle panel plots the gas pressure (solid line) and magnetic
pressure. The top right panel plots the rotational, radial,
Keplerian, and Alfv${\acute{\rm e}}$n velocities  
(solid, dashed, dot-dashed, and dotted line, respectively), as well as
the sound speed (triple-dot dashed line).
The bottom middle panel plots the Maxwell stress,  
$\alpha_{mag}=2B_rB_\phi/B^2/P_{mag}$,
and the Reynolds stress, $\alpha_{gas}=<\rho v_r \delta v_\phi>/P$ 
(solid and dashed line, respectively). 
We calculate the Reynolds stress using eq. 15 and  show only its amplitude.
The bottom right panel plots
the radial, latitudinal and toroidal components of the magnetic field
(dot-dashed, dashed, and solid line, respectively).

Fig. 10. As Fig. 8 but for run~I.

\newpage

\begin{figure}
\begin{picture}(180,580)
\put(340,330){\includegraphics{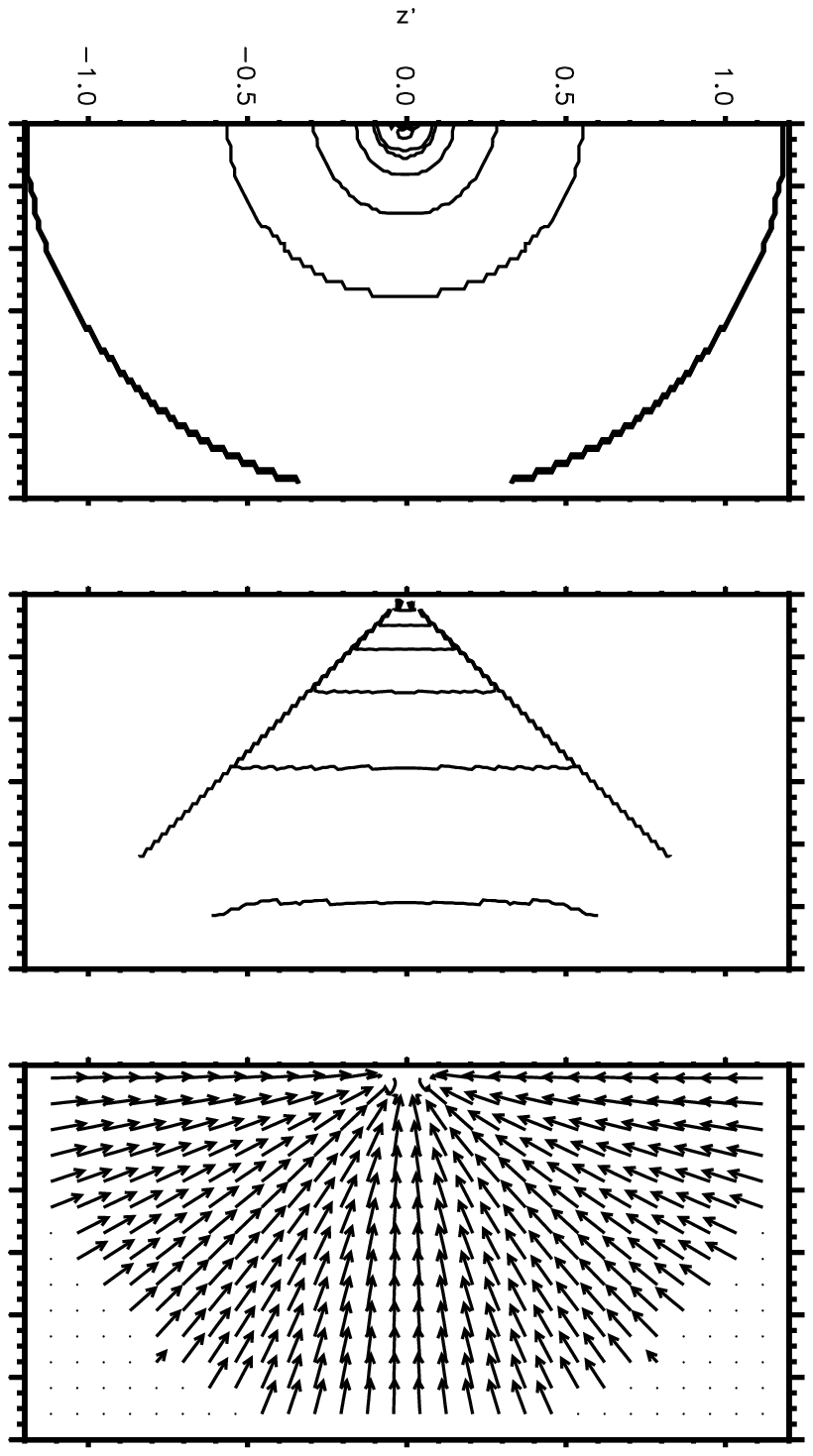}}

\put(340,210){\includegraphics{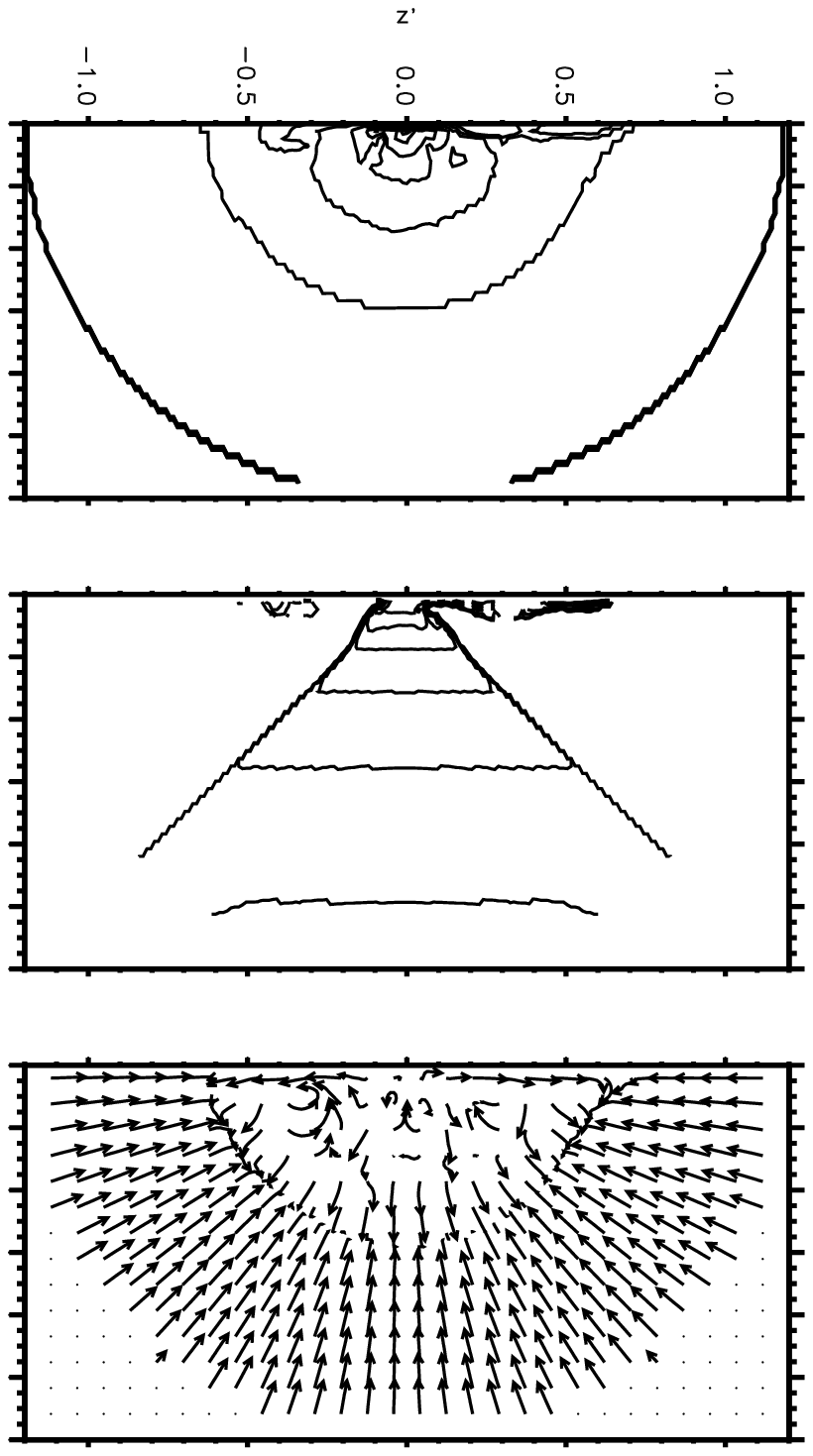}}

\put(340,90){\includegraphics{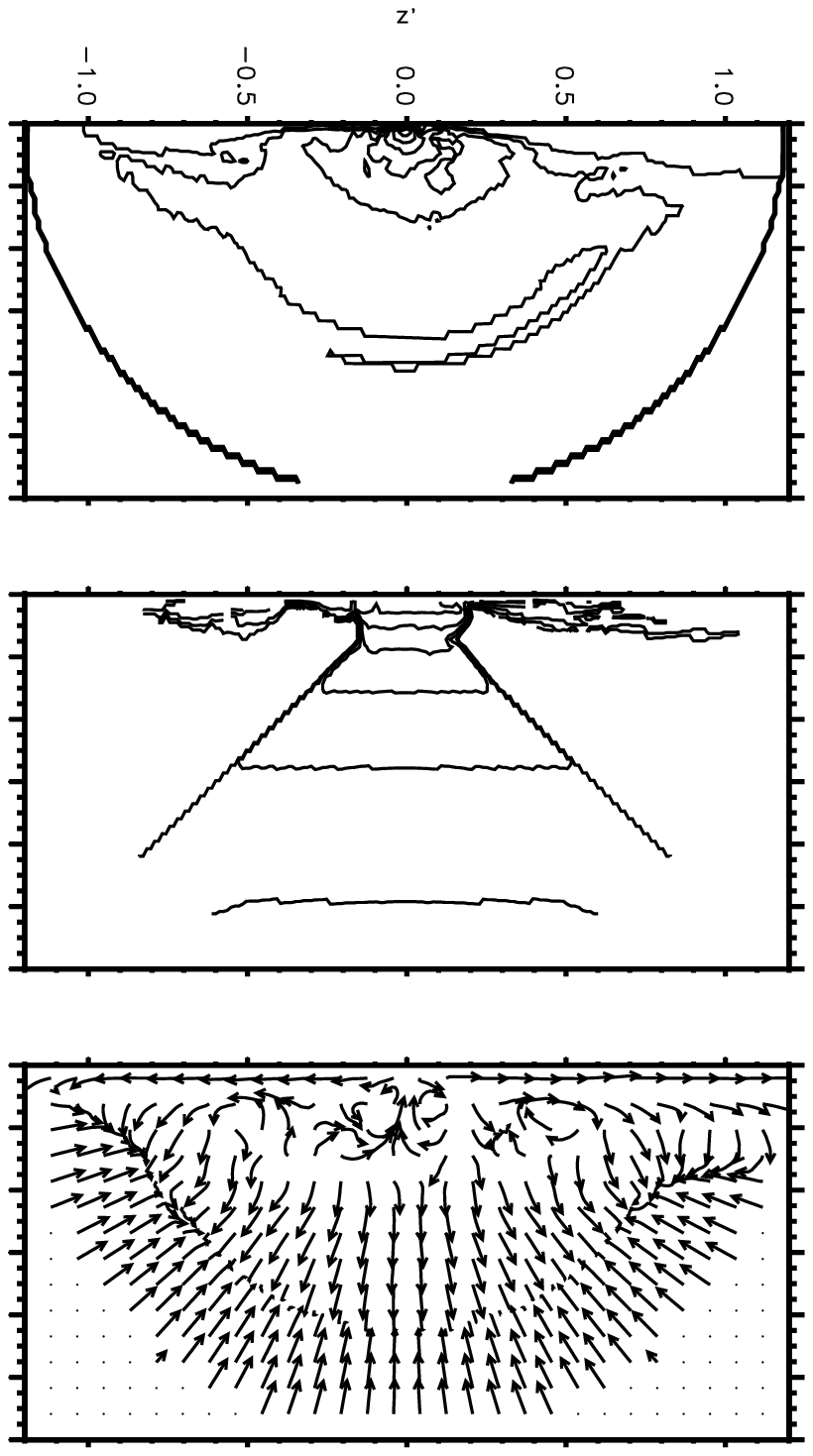}}

\put(340,-30){\includegraphics{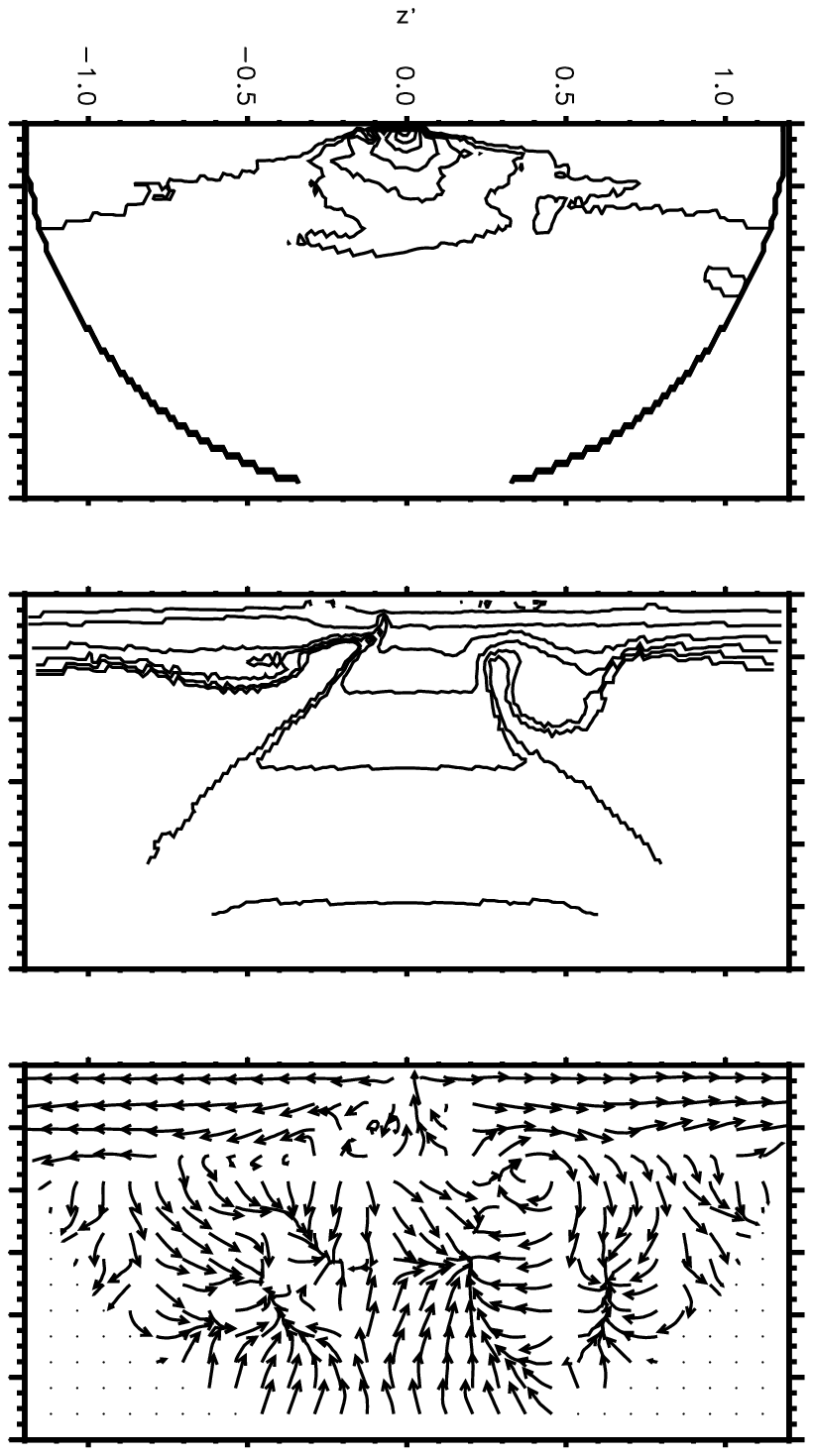}}

\put(340,-150){\includegraphics{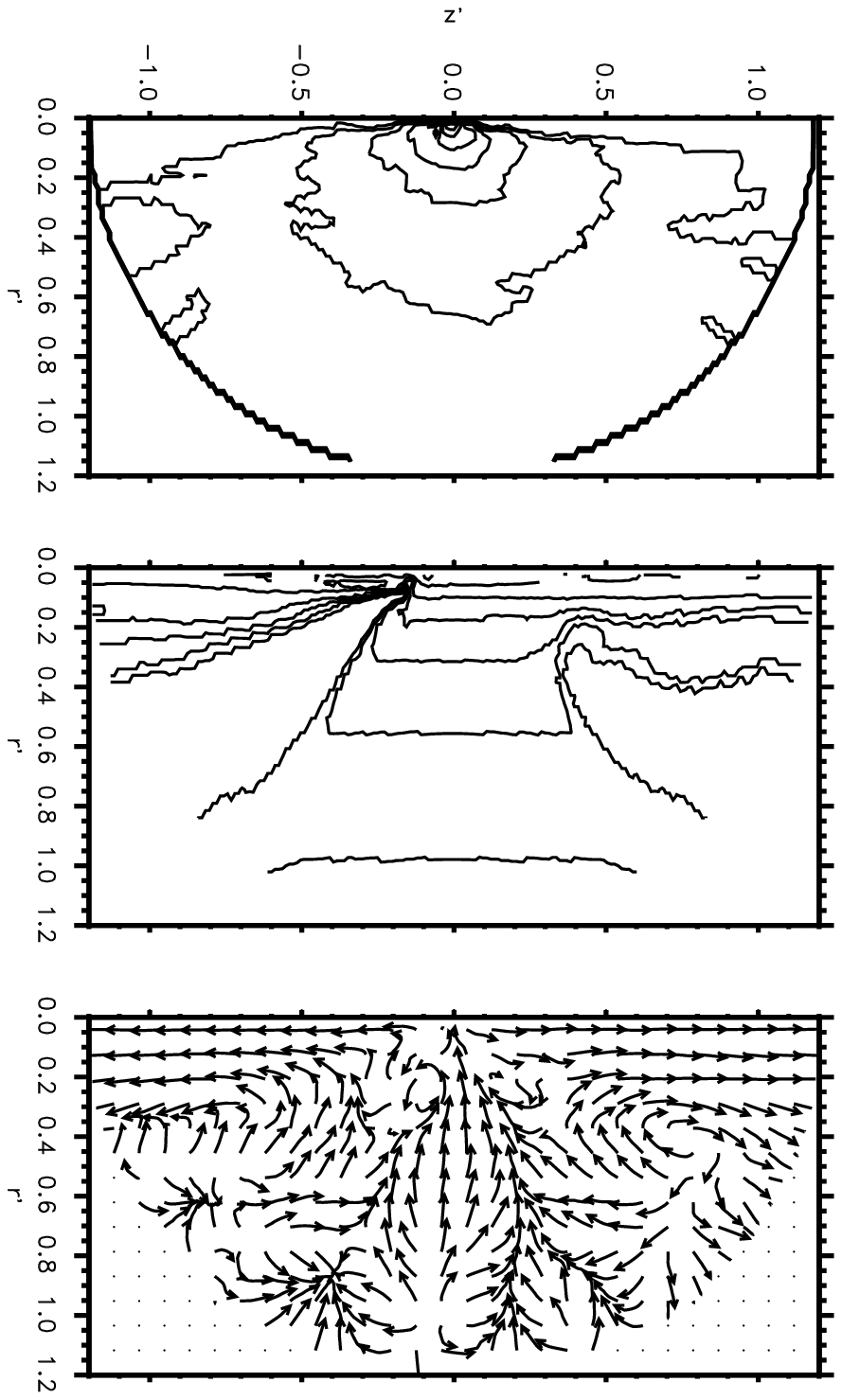}}

\end{picture}
\caption{ 
}
\end{figure}

\newpage

\begin{figure}
\begin{picture}(180,400)
\put(0,0){\includegraphics{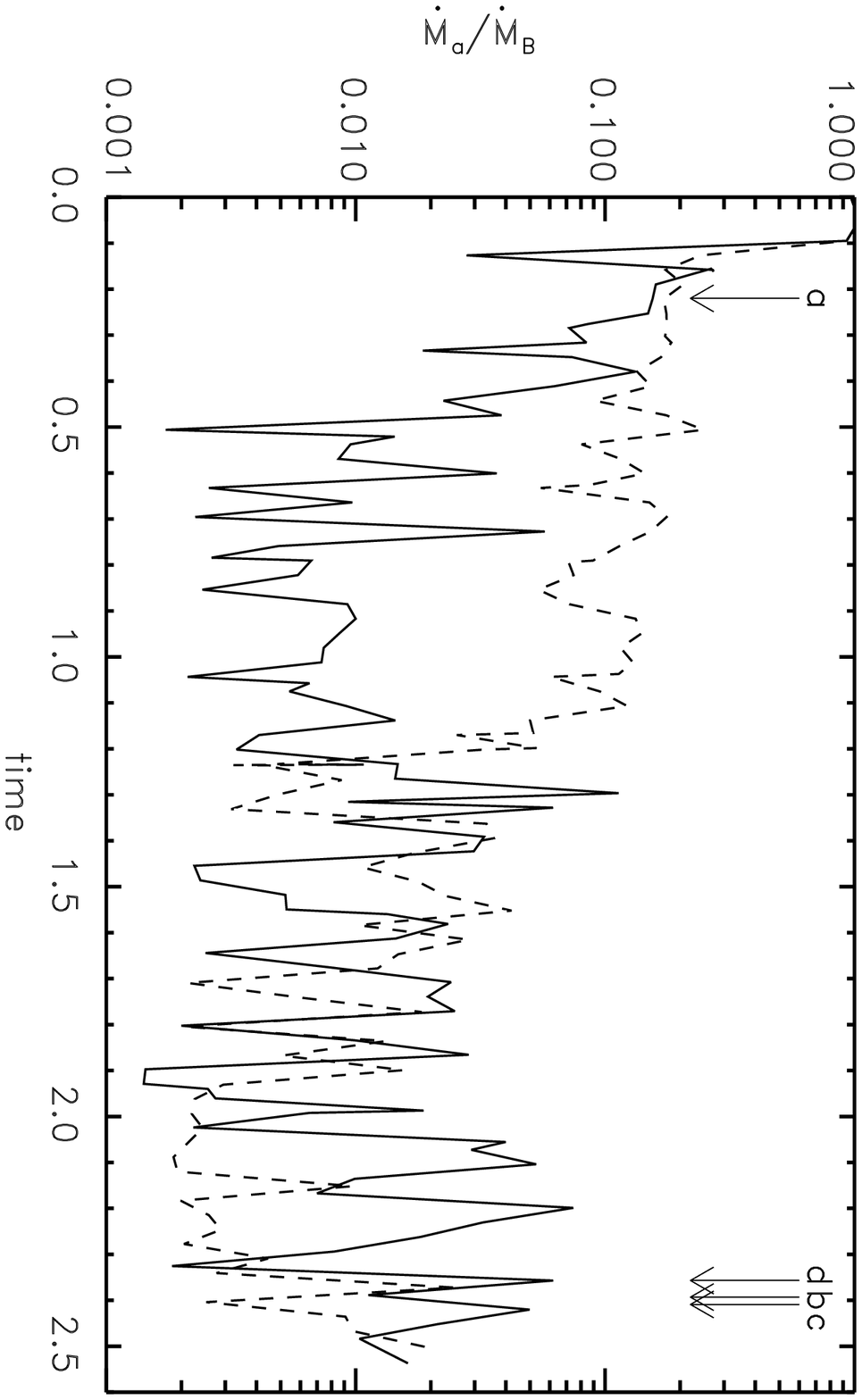}}
\end{picture}
\caption{ 
}
\end{figure}

\begin{figure}
\begin{picture}(180,500)
\put(0,200){\includegraphics{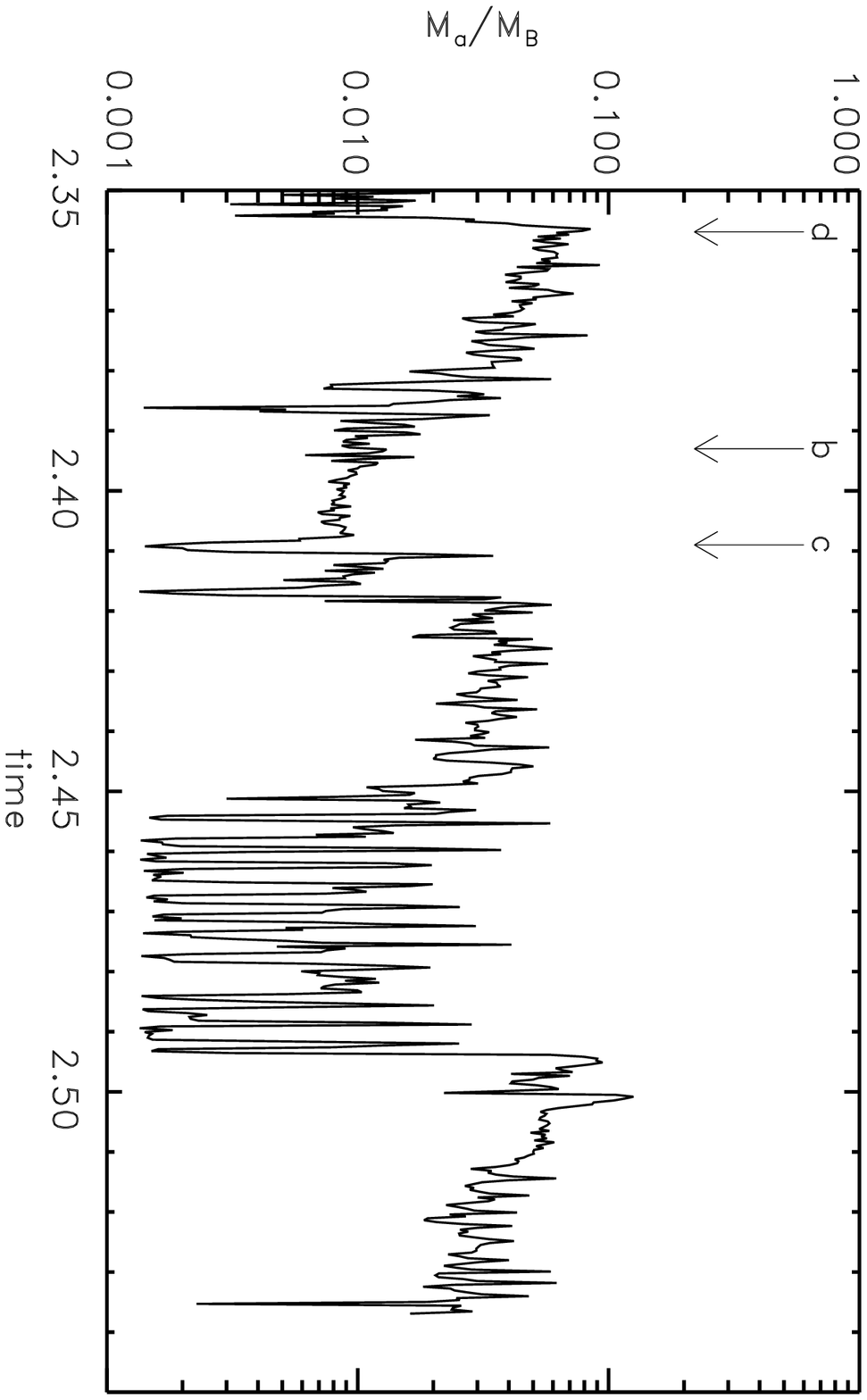}}

\put(0,-50){\includegraphics{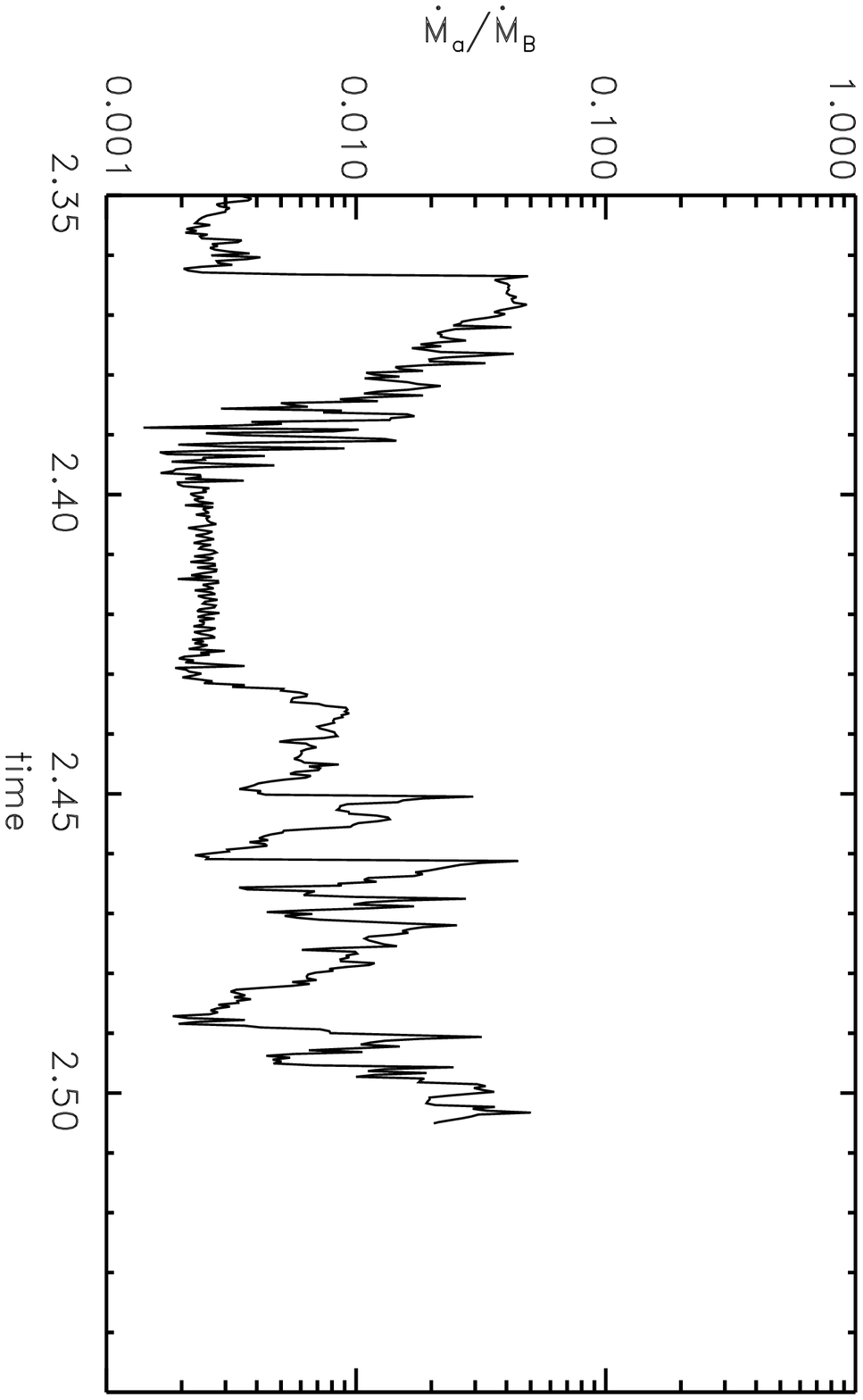}}

\end{picture}
\caption{ 
}
\end{figure}

\begin{figure}
\begin{picture}(180,400)
\put(0,0){\includegraphics{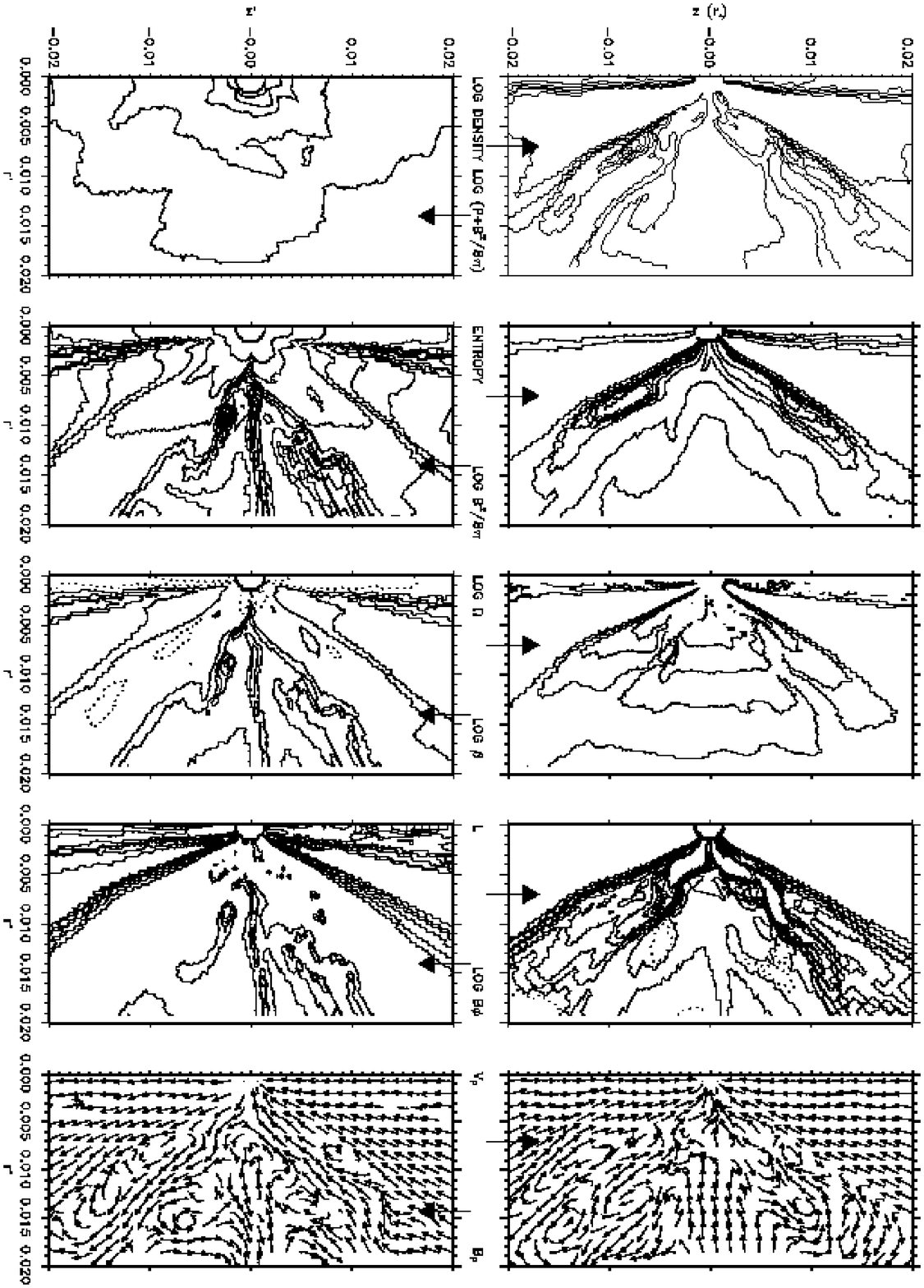}}
\end{picture}
\caption{ 
}
\end{figure}

\begin{figure}
\begin{picture}(180,400)
\put(0,0){\includegraphics{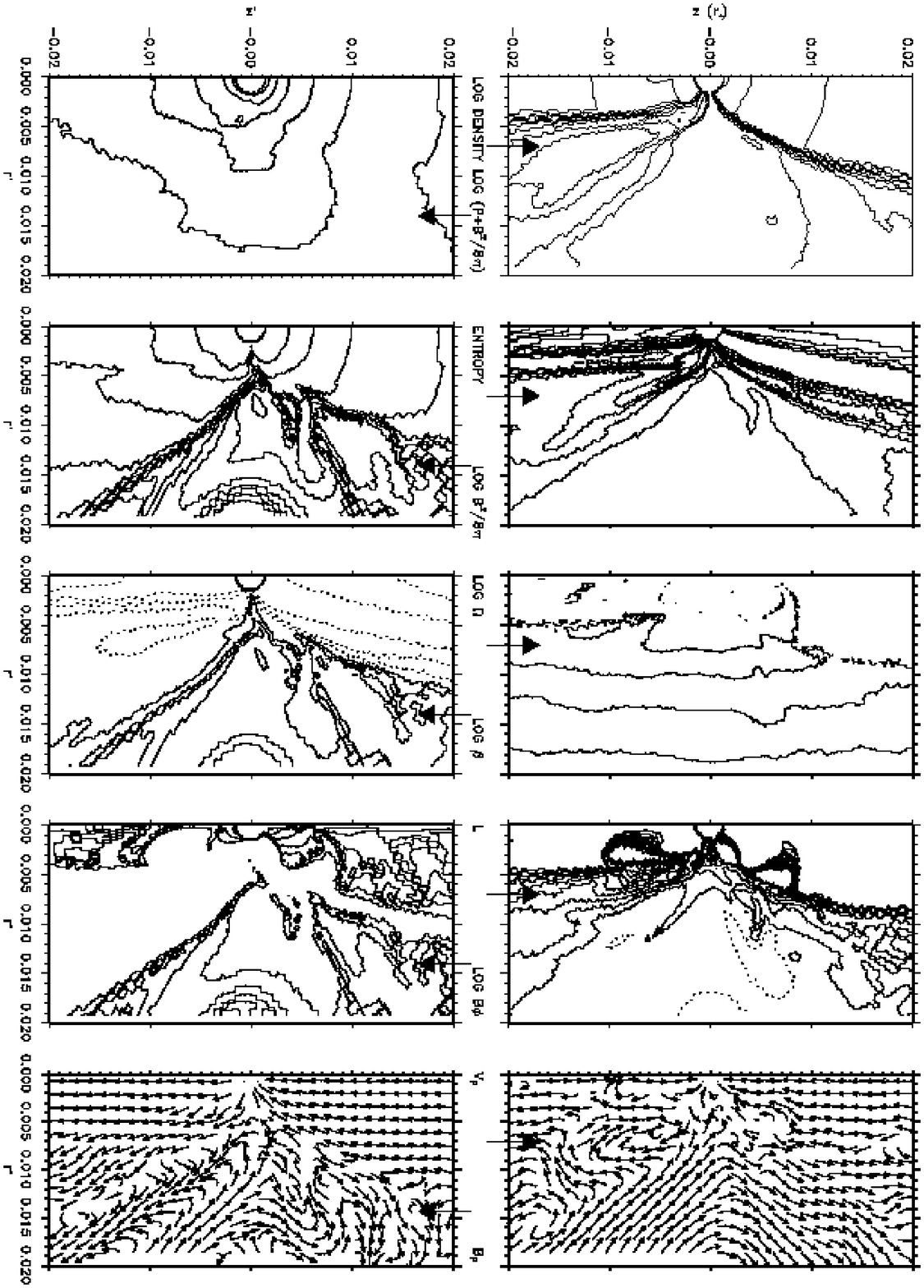}}
\end{picture}
\caption{ 
}
\end{figure}

\begin{figure}
\begin{picture}(180,400)
\put(0,0){\includegraphics{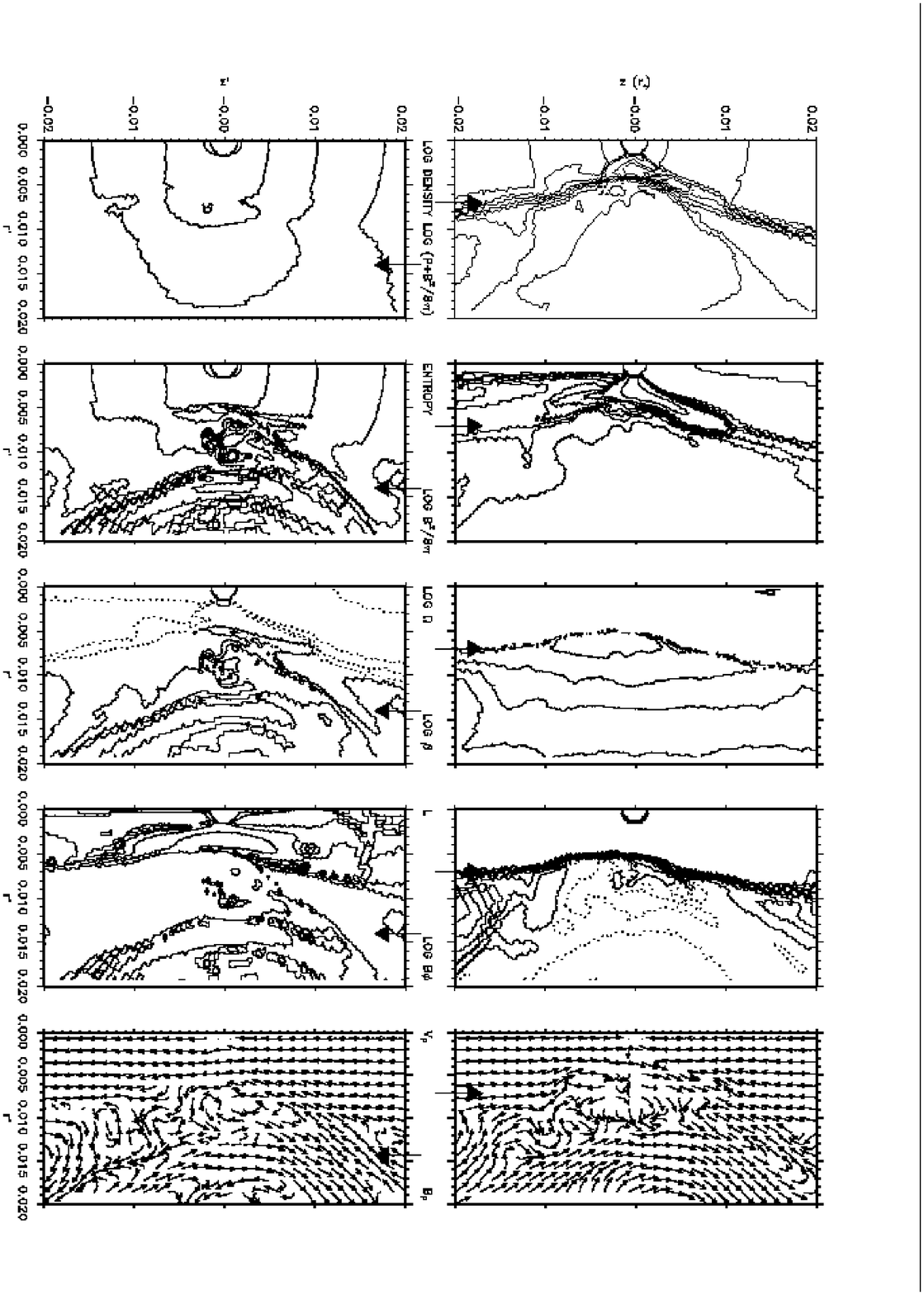}}
\end{picture}
\caption{ 
}
\end{figure}

\begin{figure}
\begin{picture}(180,400)
\put(0,0){\includegraphics{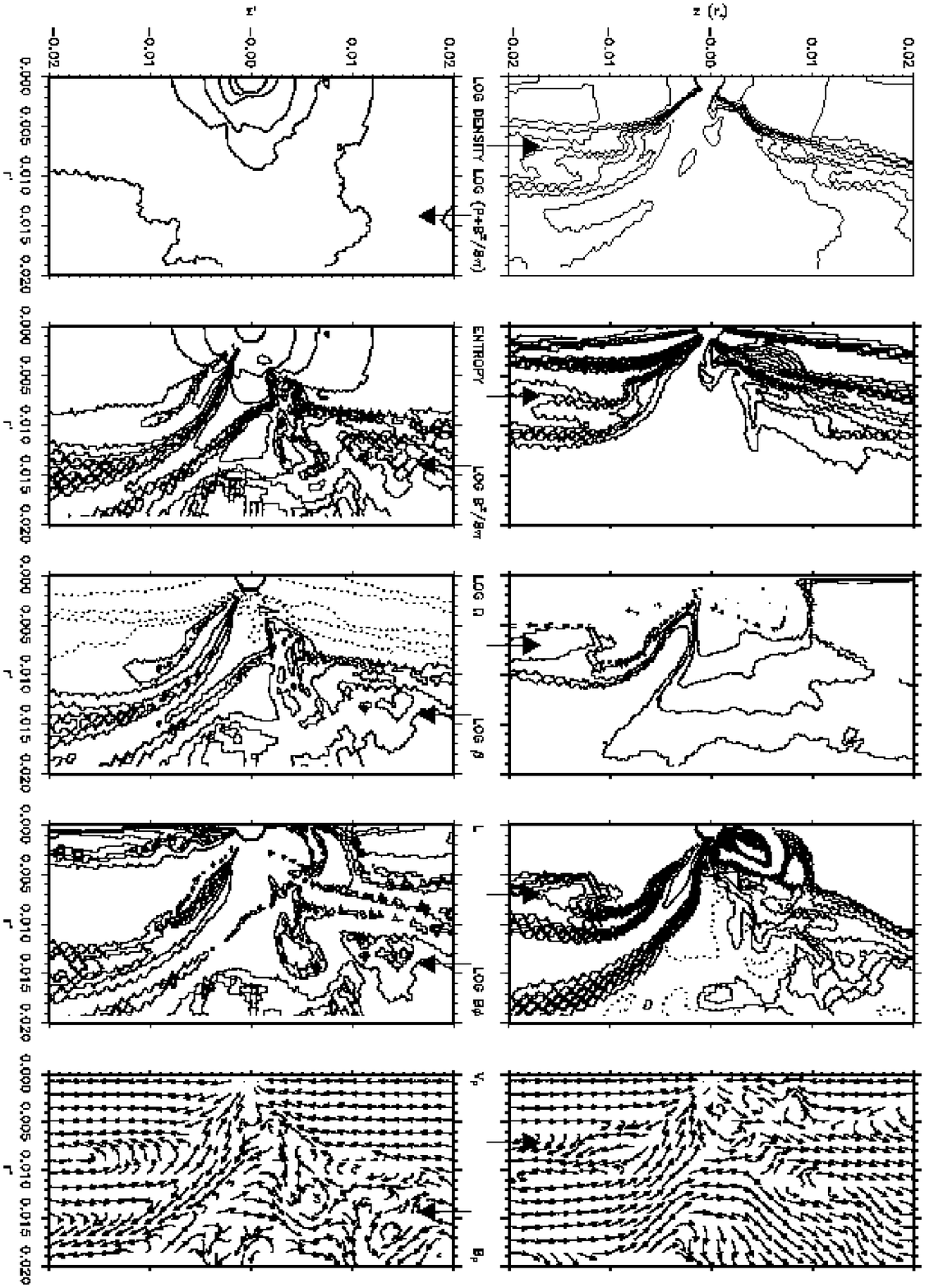}}
\end{picture}
\caption{ 
}
\end{figure}

\begin{figure}
\begin{picture}(180,560)
\put(300,300){\includegraphics{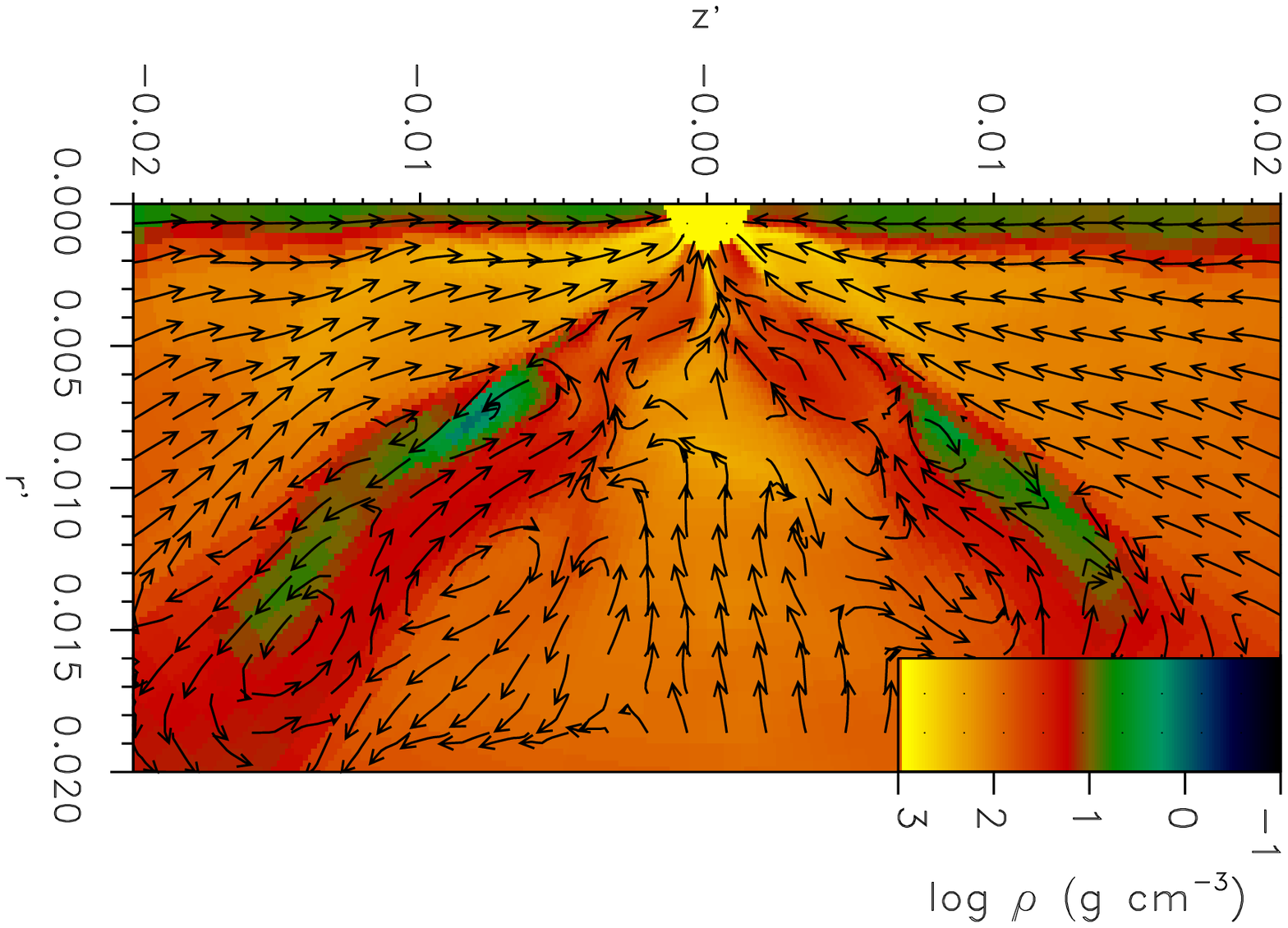}}
\put(550,300){\includegraphics{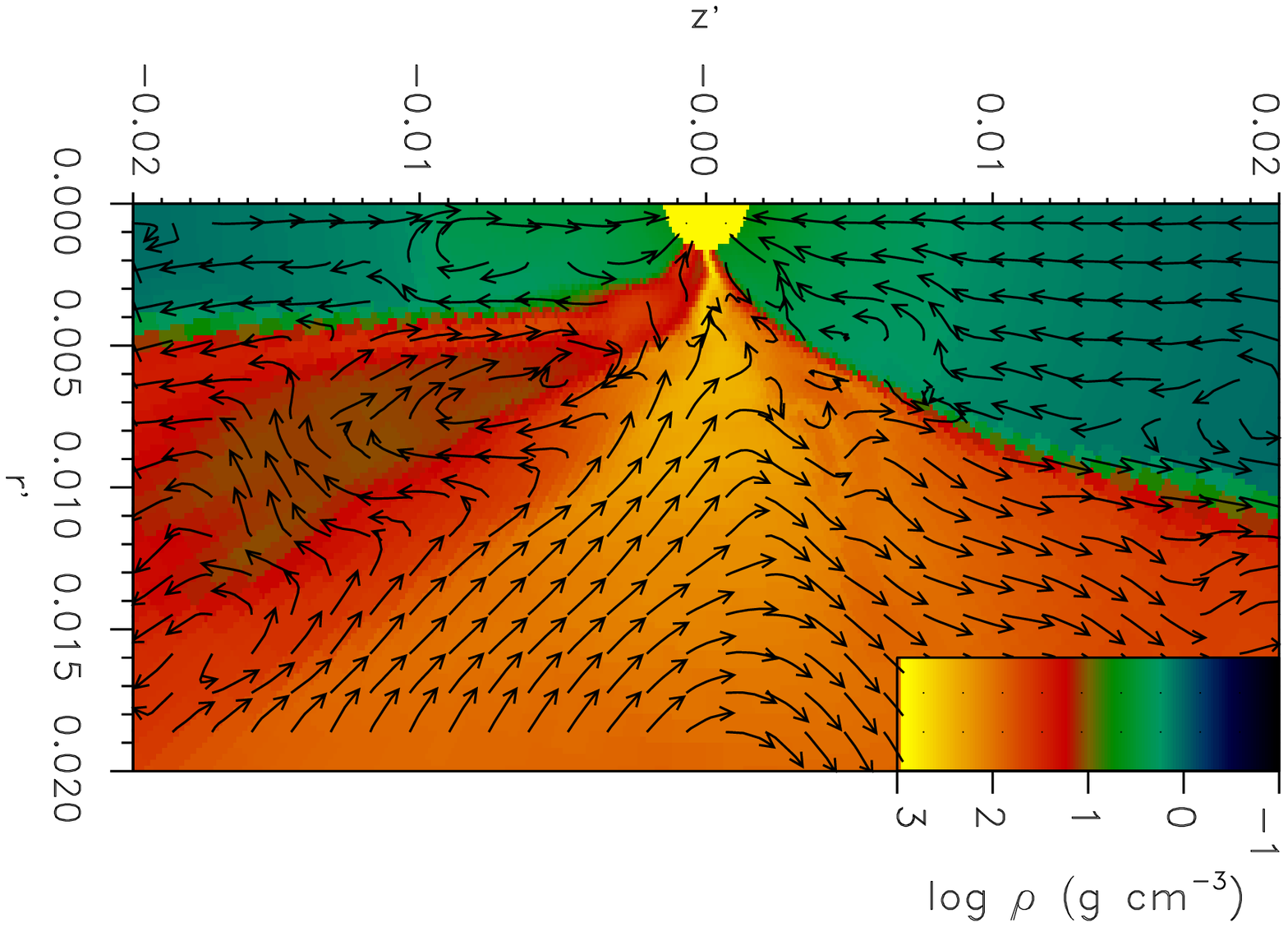}}
\put(300,0){\includegraphics{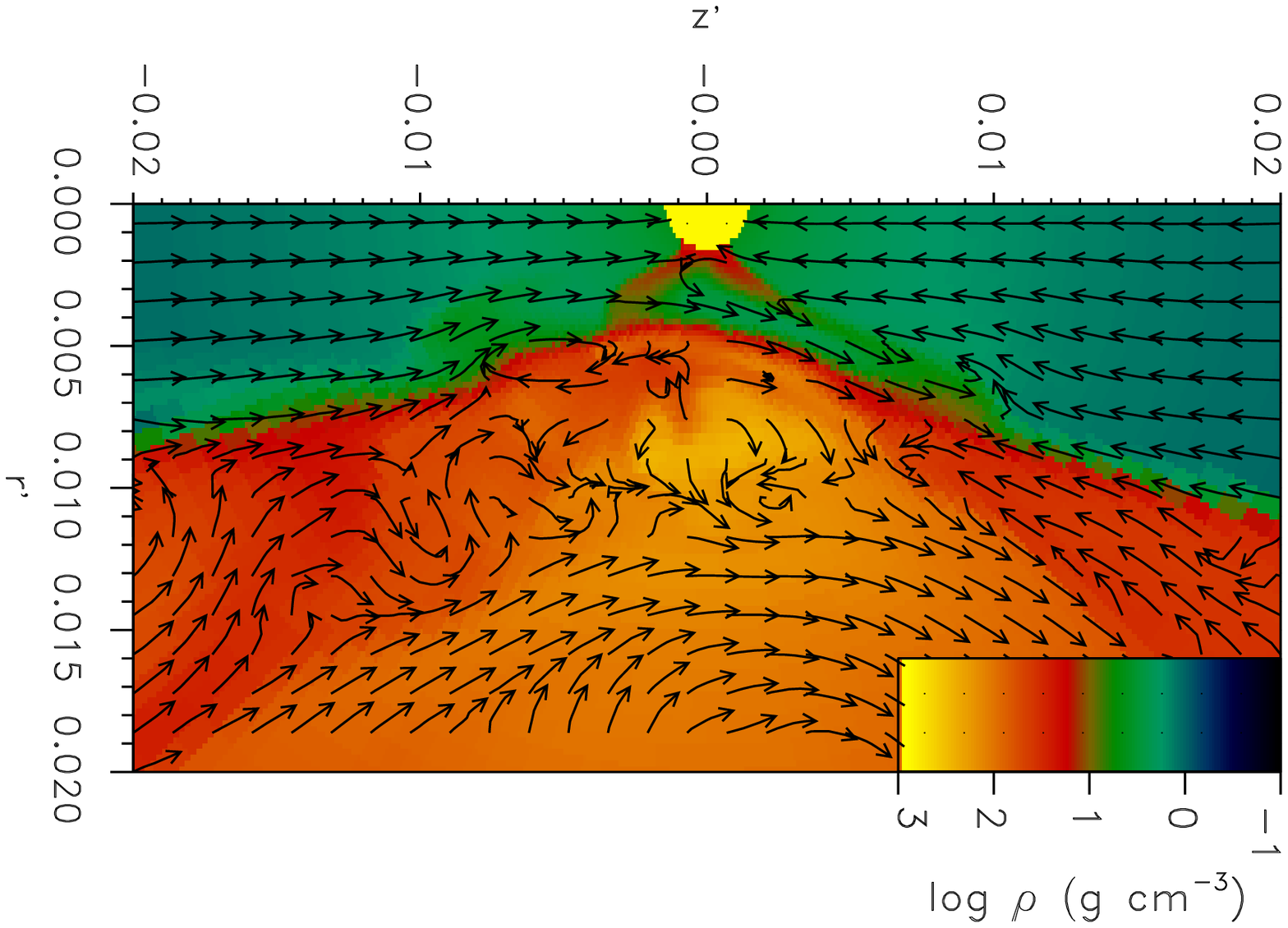}}
\put(550,0){\includegraphics{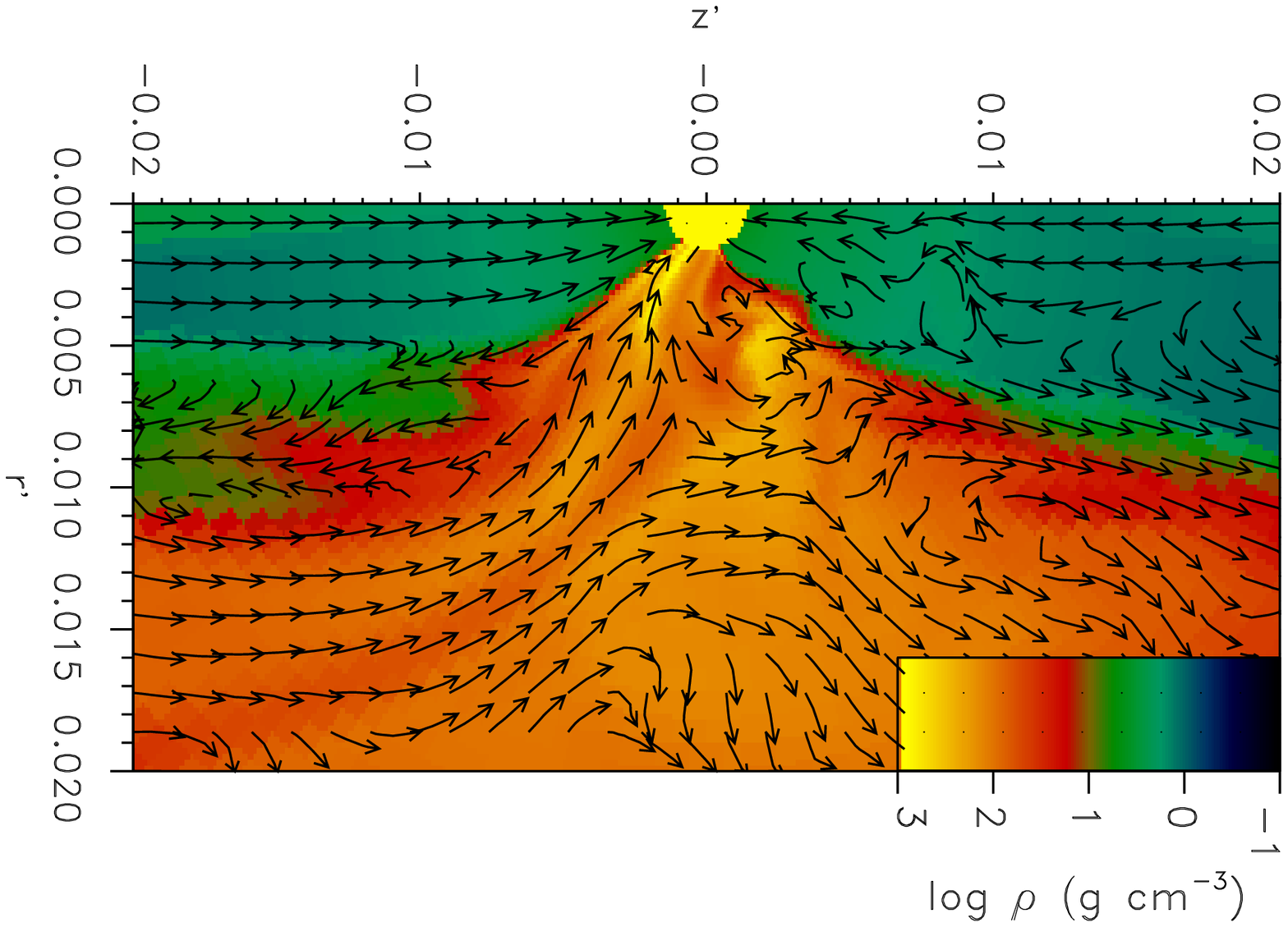}}
\end{picture}
\caption{
}
\end{figure}

\begin{figure}
\begin{picture}(180,400)
\put(0,0){\includegraphics{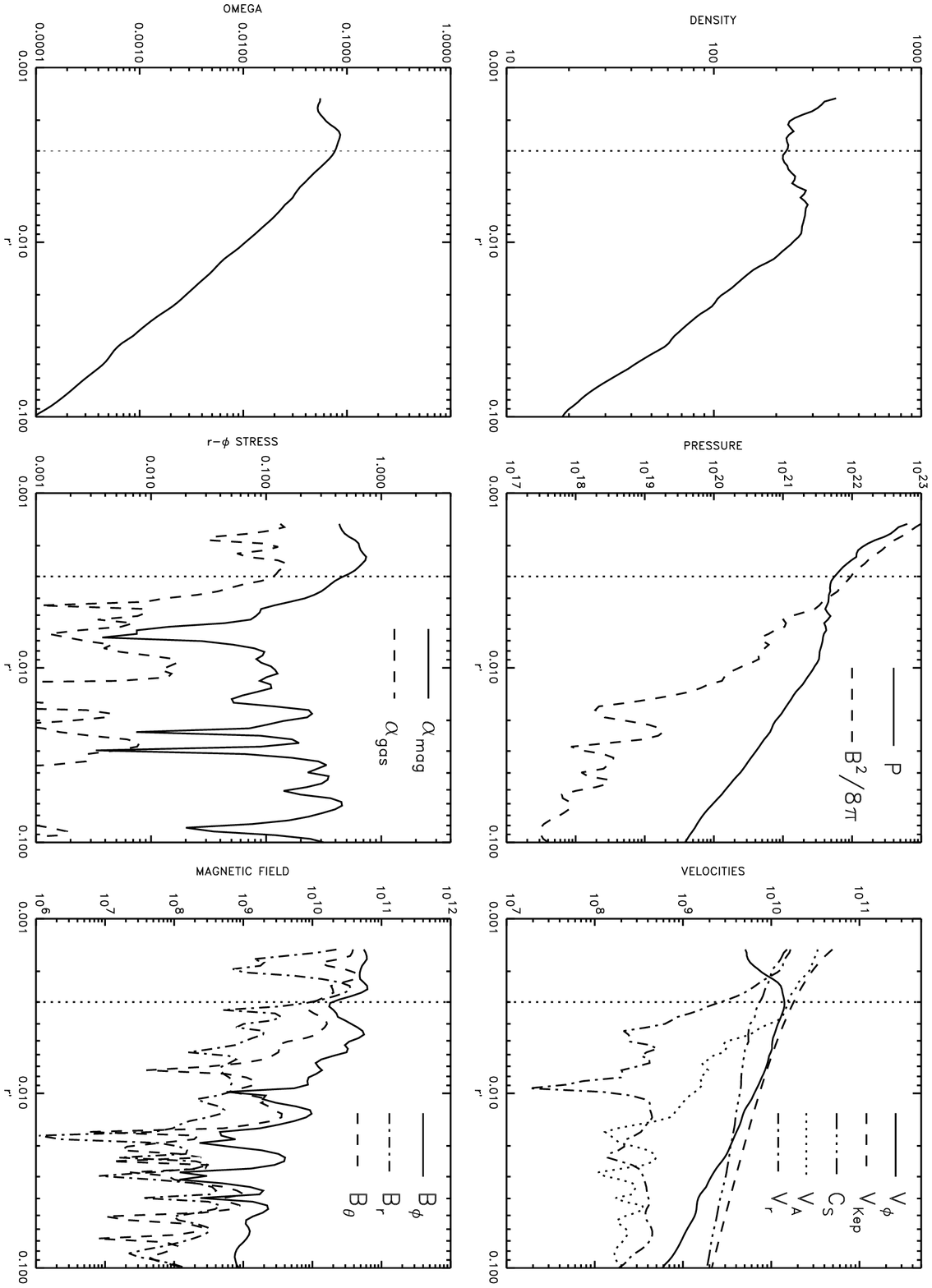}}
\end{picture}
\caption{ 
}
\end{figure}

\begin{figure}
\begin{picture}(180,400)
\put(0,0){\includegraphics{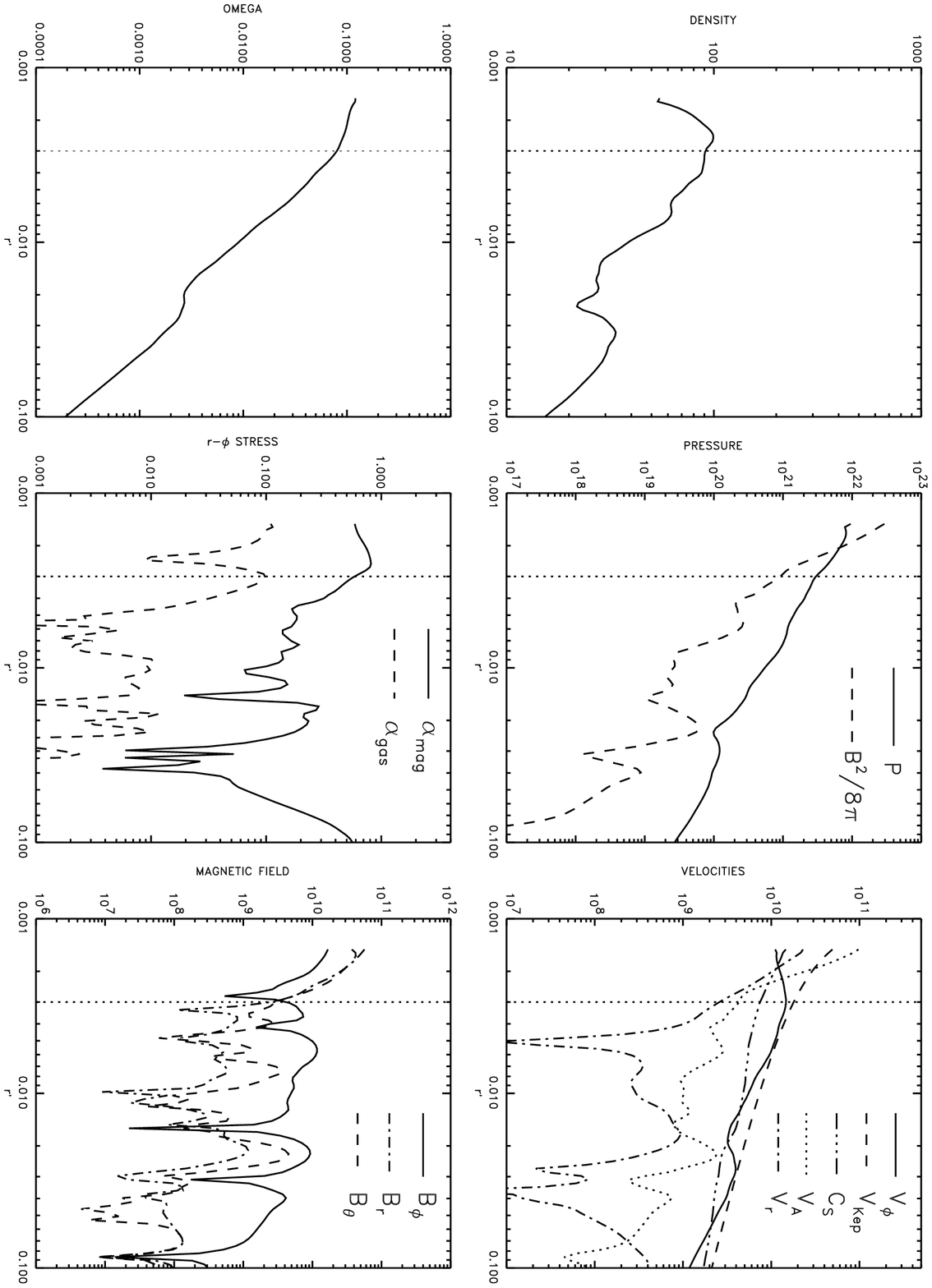}}
\end{picture}
\caption{ 
}
\end{figure}

\end{document}